%% file: main.tex
\documentclass{aa}
\usepackage[utf8]{inputenc}
\usepackage{amsmath}
\usepackage{amssymb}
\usepackage{graphicx} 
\usepackage{epstopdf}
\usepackage{arydshln}
\usepackage{txfonts}
\usepackage{multicol}
\usepackage{multirow} 
\input{bibdefinitions.tex}

\titlerunning{Phase-resolved spectroscopic analysis of the eclipsing black hole X-ray binary M33\,X-7}

\title{Phase-resolved spectroscopic analysis of the eclipsing black hole X-ray binary M33\,X-7: System properties, accretion, and evolution\thanks{Based on observations made with the NASA/ESA {\em Hubble} Space Telescope ({\em HST}), obtained from the data archive at the Space Telescope Science Institute. STScI is operated by the Association of Universities for Research 
in Astronomy, Inc. under NASA contract NAS 5-26555. These observations are associated with the GO program 15636.
Based on observations obtained with {\em XMM-Newton}, an ESA science mission with instruments and contributions 
directly funded by ESA Member States and NASA, these observations are associated with id numbers 0831590201 \& 0831590401.}}

 \author{V. Ramachandran\inst{1,2}
          \and L. M. Oskinova\inst{2}
           \and W.-R. Hamann\inst{2} 
          \and A. A. C. Sander\inst{1}
          \and H. Todt\inst{2} 
          \and D. Pauli\inst{2}
          \and T. Shenar\inst{3}
          \and J. M. Torrej\'{o}n\inst{4}
          \and K. A. Postnov\inst{5}
          \and J. M. Blondin\inst{6}
          \and E. Bozzo\inst{7}
          \and R. Hainich\inst{2}
          \and D. Massa\inst{8}
          }

   \institute{Zentrum f{\"u}r Astronomie der Universit{\"a}t Heidelberg,
Astronomisches Rechen-Institut, M{\"o}nchhofstr. 12-14, 69120 Heidelberg\\
              \email{vramachandran@uni-heidelberg.de}              
\and Institut f{\"u}r Physik und Astronomie, Universit{\"a}t Potsdam, Karl-Liebknecht-Str. 24/25, D-14476 Potsdam, Germany  
\and Anton Pannekoek Institute for Astronomy, University of Amsterdam, 1090 GE Amsterdam, The Netherlands
\and Instituto Universitario de F\'{i}sca Aplicada a las Ciencias y las Tecnolog\'{i}as, Universidad de Alicante, E-03690 Alicante, Spain
\and  Sternberg Astronomical Institute, M.V. Lomonosov Moscow University, Universitetskij pr. 13, 119234 Moscow, Russia
\and Department of Physics, North Carolina State University, Raleigh, NC 27695-8202, USA
\and Department of Astronomy, University of Geneva, Chemin d’Ecogia 16, CH-1290 Versoix, Switzerland
\and Space Science Institute, 4750 Walnut Street, Suite 205, Boulder, CO 80301, USA}       
\date{}

\abstract{
M33\,X-7 is the only known eclipsing black hole high mass X-ray binary. The system is reported to contain a very massive O supergiant donor and a massive black hole in a short orbit. The high X-ray luminosity and its location in the metal-poor galaxy M33 make it a unique laboratory for studying the winds of metal-poor donor stars with black hole companions and it helps us to understand the potential progenitors of black hole mergers.
Using phase-resolved simultaneous \hst- and \xmm-observations, we traced the interaction of the stellar wind with the black hole. We observed a strong Hatchett-McCray effect in M33\,X-7 for the full range of wind velocities. Our comprehensive spectroscopic investigation of the donor star (X-ray+UV+optical) yields new stellar and wind parameters for the system that differ significantly from previous estimates.
In particular, the masses of the components are considerably reduced to $\approx 38\,M_{\odot}$ for the O-star donor and $\approx 11.4\,M_{\odot}$ for the black hole. The O giant is overfilling its Roche lobe and shows surface He enrichment.
The donor shows a densely clumped wind with a mass-loss rate that matches theoretical predictions. An extended ionization zone is even present during the eclipse due to scattered X-ray photons. The X-ray ionization zone extends close to the photosphere of the donor during inferior conjunction.
We investigated the wind-driving contributions from different ions and the changes in the ionization structure due to X-ray illumination. Toward the black hole, the wind is strongly quenched due to strong X-ray illumination.
For this system, the standard wind-fed accretion scenario alone cannot explain the observed X-ray luminosity, pointing toward an additional mass overflow, which is in line with our acceleration calculations. 
The X-ray photoionization creates an \heii\ emission region around the system emitting $\sim 10^{47} \mathrm{ph\,s^{-1} } $. We computed binary evolutionary tracks for the system using MESA. Currently, the system is transitioning toward an unstable mass transfer phase, possibly resulting in a common envelope of the black hole and the O-star donor. Since the mass ratio is q$\gtrsim$3.3 and the period is short, the system is unlikely to survive the common envelope, but will rather merge.
}

\begin{document}

\maketitle

\section{Introduction}

\revise{The majority of massive stars are commonly found in binary or higher-order multiple systems \citep[e.g.,][]{sana2012,Sana2014}. The evolution of these massive stars is largely governed by the loss of mass via stellar winds, and by the exchange of mass among binary components \citep[e.g.,][]{langer2012,Postnov2006}. 
When the initially more massive (i.e. primary) component in a binary system collapses into either a neutron star (NS) or a black hole (BH) at the end of its evolution, some systems remain bound after the supernova explosion. The accretion of mass transferred from the former secondary star onto its compact companion can power strong X-ray emission \citep{Shklovsky1967}. Such systems are called high-mass X-ray binaries (HMXBs). The majority of HMXBs contain NSs, while systems containing BHs are quite rare. In this paper, we focus on M33\,X-7, an HMXB with a BH companion.}

Black hole X-ray binaries (BH XRBs) are essential natural laboratories for studying stellar-mass BHs. These systems provide an indispensable observational test for BHs that are formed by core collapse, therefore providing observational constraints for binary evolution and compact object formation. The majority of the detected BH XRBs have low-mass donor stars and are X-ray transients \citep{Remillard2006,Ozel2010,Corral-Santana2016}. BH XRBs with massive donor stars are persistently strong X-ray sources that are powered by massive stellar winds and/or Roche lobe overflow. Cyg X-1 was the first discovered BH HMXB system (BH + O9I companion) and it remains the only confirmed one in the Milky Way \citep[e.g.,][]{Paczynski1974,GiesBolton1986,Miller-Jones+2021}. 
Other BH HMXB candidates in the Galaxy are SS433 \citep{Cherepashchuk2020}, Cyg X-3 \citep{Zdziarski2013}, and  MWC656 \citep{Casares2014}.
In the extra-galactic population of BH HMXBs, dynamical evidence has been presented for LMC\,X-1 \citep[BH + O7\,III companion;][]{Orosz2009}, LMC X-3 \citep[BH + B3–5,\,V star][]{Orosz2014}, and M33\,X-7 \citep[BH with an O7-8\,III companion;][]{Orosz2007}. There is indirect evidence for the presence of a BH with a Wolf-Rayet companion in NGC300\,X-1 \citep{Binder2021,Crowther2010NGC300} and IC10\,X-1 \citep{Laycock2015,Silverman2008} systems.

If the donor star in a BH HMXB is massive enough, its core collapse can result in the formation of a relativistic binary comprising two BHs. In the case of a tight system, the binary components may merge within a Hubble time. Such a merger may be observed as a gravitational wave (GW) event \citep[e.g.,][]{Abbott+2016a,Abbott+2016b,Abbott+2017}. Thus, HMXBs with BHs represent a key transitional stage toward a binary BH merger. It is crucial to understand the physics of these systems and their dependence on metallicity in order to establish whether the BHs in present-day massive binaries and those detected by GWs during a merger are part of the same stellar progenitor pool.

In HMXBs, accretion onto the compact object is typically powered by the stellar wind of the donor star.  For  Bondi-Hoyle-Lyttleton accretion \citep{Bondi1944,Davidson1973,Edgar2004}, the X-ray luminosity crucially depends on the wind parameters of the donor star. In particular, Bondi-Hoyle mass accretion is highly dependent on the velocity at which the donor wind is captured by the gravitational field of the compact object  \citep[$\varv^{-3}$, from][]{Bondi1944}. The X-ray luminosity is directly dependent on the accretion rate and accretion mode and can vary rapidly due to the complexity of the stellar wind and the possible clumping of material. Hence it is of the utmost importance to understand the wind properties of the massive donor. Sophisticated spectral analyses of the donor stars using state-of-the-art model atmospheres to determine the stellar and wind parameters of the donor stars in HMXBs are therefore a pivotal instrument for a better understanding of these systems.

The winds of OB stars are driven by the scattering of UV photons in metal lines \citep{Castor1975}. At lower metallicity, the winds of the OB stars are much weaker \citep{bouret2015,ramachandran2019} and hence remove less mass and angular momentum during stellar evolution. Thus, a metal-poor star is expected to undergo its core collapse at a higher mass and at faster rotation than a star with initially the same mass, but higher metallicity. As of yet, virtually nothing is known about the winds of metal-poor donor stars in HMXBs with BH companions. \revise{Our current understanding of the wind of massive stars in BH binaries is mostly based on the Cygnus X-1 system in our Milky Way. Yet, this prototypical system has not been analyzed with the current generation of expanding stellar atmosphere codes. Spectroscopic analyses of the BH HMXBs were limited to the Sobolev with Exact Integration (SEI) method \citep[e.g., Cyg X-1 in][]{Gies2008,Vrtilek2008} or using plane-parallel models (e.g., for M33 X-7 by \citealt{Orosz2007} and for Cyg X-1 by \citealt{Caballero-Nieves2009}) which are optimized for hot stars with no significant wind.}
More advanced analyses of donors done in the recent past are limited to NS HMXB systems. For example, a sample of HMXBs with NS companions was recently studied by \cite{Hainich2020} with detailed expanding stellar atmosphere models for radiation-driven winds. 
In this study, we carry out a detailed atmospheric analysis of multiwavelength spectra of the metal-poor donor star in M33\,X-7, taking into account line-driven winds, wind clumping, complex effects of millions of spectral lines, and X-ray photoionization.

\begin{table}
\caption{M33\,X-7 system parameters from the literature used in this work.}
    \label{table:systemparameters}
    \centering
    \renewcommand{\arraystretch}{1.4}
\begin{tabular}{lcc}
\hline 
\hline
\vspace{0.1cm}
Parameter                  & Value & Ref. \\
\hline
$d$ [kpc]               &  840 $\pm$ 40  & 1,2,3   \\
$m_{\mathrm{v}}$ [mag]              &  18.9   & 1 \\
$L_{\mathrm{x}}$ (0.5-5\,kev) [erg\,s$^{-1}$]  &  $\sim$(0.02-2) $\times 10^{38}$    & 1,4,5  \\
$T_{\mathrm{0}}$ [HJD]                & $2,453,967.157\pm0.05$     &  1\\
$P_{\mathrm{orb}}$ [days]              &  3.45301 & 1,4 \\
systemic velocity [km\,s$^{-1}$]&     $152\pm5$   &1\\
eccentricity $e$     &    $0.0185\pm0.0077$   & 1\\
$f(M)$ [$M_{\odot}$]            &    $0.46\pm0.08$  &1 \\ 
\hline 
\end{tabular}
\tablebib{
(1)~\citet{Orosz2007}; (2) \citet{Galleti2004}; (3) \citet{Gieren2013}; (4) \citet{Pietsch2006}; (5) \cite{Pietsch2004}.
}
\end{table}
\medskip

\subsection*{The M33\,X-7 system}

M33\,X-7 is reported to \revise{contain one of the most massive known stellar mass BHs} with a very massive O7\,III star in a tight binary orbit \citep{Orosz2007}. The system is located in the metal-poor galaxy of Messier 33 (M33). M33 is known to have a steep metallicity gradient \citep{Cioni2009,Magrini2007}. The M33\,X-7 system is located in an OB association (HS\,13) approximately $ 2$\,kpc from the center of M\,33. Based on \hii region abundance measurements by \cite{Magrini2007}, the metallicity at this galactic radius is roughly half solar.

Importantly, M33\,X-7 is the only known eclipsing BH HMXB \citep{Pietsch2004,Pietsch2006}. \cite{Pietsch2006} derived a period of 3.45 days and an X-ray luminosity of $5-11 \times 10^{37}$\,erg\,s$^{-1}$ out of eclipse.
The study conducted by \cite{Orosz2007} provided a dynamical estimate of the mass of the system, reporting a $\approx\!70M_{\odot}$ O supergiant donor, and an $\approx\!16M_{\odot}$ BH. The mass of the donor is  very high for an O7\,III spectral type \citep{Markova2018,ramachandran_2018_stellar-2} as well as compared to O stars in other BH HMXBs. Such high  masses for the donor and compact companion are exceptional among HMXBs, making M33X-7 a unique laboratory for studying the potential progenitors of massive BH mergers detected by LIGO. 

Despite a rather tight orbit, it was reported that the O-type donor does not fill its Roche lobe radius \citep[$R_\ast \approx 0.77 R_\mathrm{rl}$, ][]{Orosz2007}, and therefore the X-ray luminosity is expected to be powered by accretion of the donor’s stellar wind alone. 
\cite{Orosz2007} estimated the donor mass-loss rate from the X-ray luminosity by adopting an ad hoc accretion efficiency. The estimated mass-loss rate turned out to be five times higher than theoretically expected. Furthermore, an ad hoc wind velocity law (and consequently the radial density structure of the wind) was adopted for the donor star. \revise{On the basis of these assumptions, the eclipse width was measured from the light curve with a suggested accuracy of 1 degree in terms of orbital phase angle.} As a result, it was claimed that the component masses are estimated with 10\% accuracy.

M33\,X-7 has been subject to evolutionary modeling in past studies \citep{Abubekerov2009,Valsecchi2010Natur,deMink2010ASPC, Bogomazov2014ARep}.
A large theoretical effort is ongoing to explain the large masses of the M33X-7 components, its tight orbit, and the large BH spin. Measuring the stellar wind parameters of M33\,X-7 will significantly constrain evolutionary models and allow unique tests of the evolution of BH binaries at low metallicity.

For this study, we have secured  quasi-simultaneous multiwavelength spectroscopy of M33\,X-7 at three key binary orbital phases (see Fig.\,\ref{fig:orbit}). Using the combined analysis of the multiwavelength data (UV, optical, and X-rays) we measured the properties of the donor star wind and the mass accretion rate onto the BH, and we study the effect of X-rays on the donor star wind. These wind parameter measurements are required to understand how much wind is quenched by the X-rays, so as to test stellar wind theory, refine the measurements of the BH mass, and determine the accretion efficiency onto the BH. Subsequently, our new realistic wind parameters serve as input for sophisticated massive binary evolution models. The basic parameters of the system which are used in this study from the abovementioned literature are listed in  Table.\,\ref{table:systemparameters}. Throughout the paper, we assume a distance of 840\,kpc to M33 \citep{Galleti2004,Gieren2013}.

This paper is organized as follows: We discuss the new and archival multiwavelength spectroscopic observations of M33\,X-7 in Sect.\,2. Based on the phase-resolved UV spectroscopy, we report orbital variations in the wind line profiles in Sect.\,3.  In Sect.\,4, we describe the detailed analysis of the spectra. In Sect.\,5 we present the updated stellar parameters of the system. We address the detailed modeling of the donor star wind in Sect.\,6 and estimate the accretion rate in Sect.\,7. Section\,8 discusses the evolution and formation of the system using binary evolution models. 
Sect.\,9 contains our final conclusions about this BH HMXB system. 

\section{Observations}

\begin{table*}
\centering
\caption{Log of spectroscopic observations used in this work}
\label{tab:observation}
\renewcommand{\arraystretch}{1.4}
\begin{tabular}{lc|ccc|ccc}
\hline 
\hline
 & & \multicolumn{3}{|c}{HST}             & \multicolumn{3}{|c}{XMM}    \\
Label               & Date   & \multicolumn{1}{|c}{JD} & \multicolumn{1}{c}{Orbital phase} & \multicolumn{1}{c}{exposure} & \multicolumn{1}{|c}{JD} & \multicolumn{1}{c}{Orbital phase} & \multicolumn{1}{c}{exposure} \\
& &\multicolumn{1}{|c}{} &\multicolumn{1}{c}{$\phi$} & \multicolumn{1}{c}{[ks]} &\multicolumn{1}{|c}{} & \multicolumn{1}{c}{$\phi$} & \multicolumn{1}{c}{[ks]} \\
 \hline  
Eclipse              & 2020-09-19 & 2459111.86 & 0.92 - 0.98 & 4 orbits &     &        &          \\
Quadrature           & 2019-07-13 & 2458678.08 & 0.3 - 0.36  & 4 orbits & 2458678.21    & 0.33-0.39   & 20\,ks   \\
Inferior Conjunction & 2019-07-17 & 2458682.25 & 0.5 - 0.56  & 4 orbits & 2458682.16    & 0.47-0.53   & 15\,ks  \\
\hline
\end{tabular}
\end{table*}

\begin{figure}
\centering
\includegraphics[width=0.47\textwidth,trim={2cm 0cm 0 -1cm}]{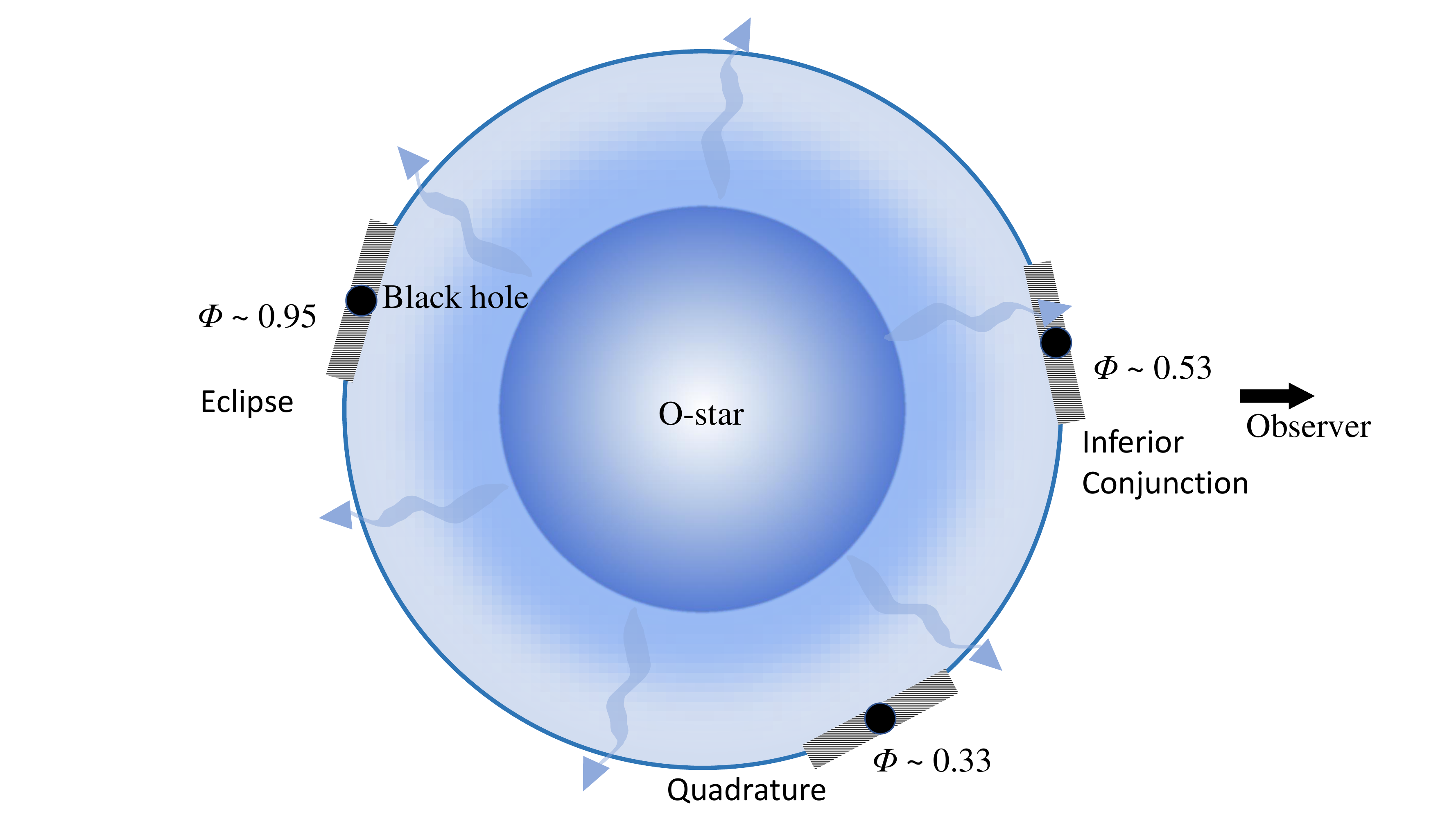}
\caption{Sketch of  the binary orbit in M33\,X-7. The observer is
located along the horizontal axis on the right. The thick lines span the range of
orbital phases covered by the \hst\ observations. The orbital separation is taken as $\approx1.75$ stellar radius as obtained in this work (see Table\,\ref{table:parameters}). }
\label{fig:orbit}
\end{figure}

We carried out simultaneous \hst\ and \xmm\ observations of M33\,X-7. These includes observations at two different phases of the orbit: quadrature ($\phi\approx0.33$)  and inferior conjunction ($\phi\approx0.53$). These two observations were taken four days apart in July 2019.
In addition we acquired \hst\ spectra during X-ray eclipse ($\phi\approx0.95$) of the system in September 2020. From the latter observations we want to obtain stellar and wind parameters of the O-star largely undisturbed by the BH. The details of the spectroscopic observations are summarized in Table\,\ref{tab:observation}.
A schematic illustration of orbit with the observed phase ranges is shown in Fig.\,\ref{fig:orbit}.

The planning of the joint {\em HST} and {\em XMM-Newton} campaign relied on the 
X-ray light-curve and ephemeris which were based on observations with the 
{\em Chandra} X-ray telescope in 2005 \citep{Pietsch2006}. According to \cite{Pietsch2006}, 
the mid-eclipse time  is occurring at HJD$(2453639.119 \pm 0.005) \pm N(3.453014 \pm 0.000020)$.  
However, the BH binaries are dynamic systems, therefore the ephemeris might not be stable and the BH might change states. 

To check the validity of ephemeris we obtained monitoring observations of M33\,X-7 
with the Neil Gehrels {\em Swift} Observatory.  {\em Swift} visited M33\,X-7 seven times  
between 2019-07-02 and 2019-07-08. During five visits, the count rates were consistent with that of the {\em Chandra}'s ACIS-I detections in 2005. On the other hand, on HJD\,2458666.828 and HJD\,2458670.147 the system was not detected in X-rays which we 
interpret as X-ray eclipses. Using the ephemeris in \cite{Pietsch2006}, these correspond to $N=1456.035$ 
and $N=1456.996$ cycles respectively, that is to say they are in agreement within uncertainties (the eclipse duration is $\approx 0.2$ in phase). Thus we verified that the orbital parameters of the system are still valid and hence were applied for scheduling our observations at three key orbital phases.

\subsection{UV spectroscopy using HST/COS}
\label{UV spectra observation}
We secured a total of 12 \hst\ orbits in cycle 26 under the GO program 15636 (PI: Oskinova). Each of the three selected phases (see Fig.\,\ref{fig:orbit}) was observed during four orbits. We selected the COS FUV G140L grating with a central wavelength setting of 1280 to cover the entire 912–2150 \AA\ wavelength range \citep{Green2012}. This configuration includes two spectral segments A and B with a gap in the range of 1192–1265 \AA, omitting the Ly$\alpha$. We disregard wavelengths $\lambda<1100$\,\AA\ and $>1800$\,\AA\ due to low signal-to-noise ratio. The resulting UV spectra have a spectral resolving power of 1500-2000. The observations were carried out in TIME-TAG mode.

\begin{figure}
\centering
\includegraphics[width=0.48\textwidth]{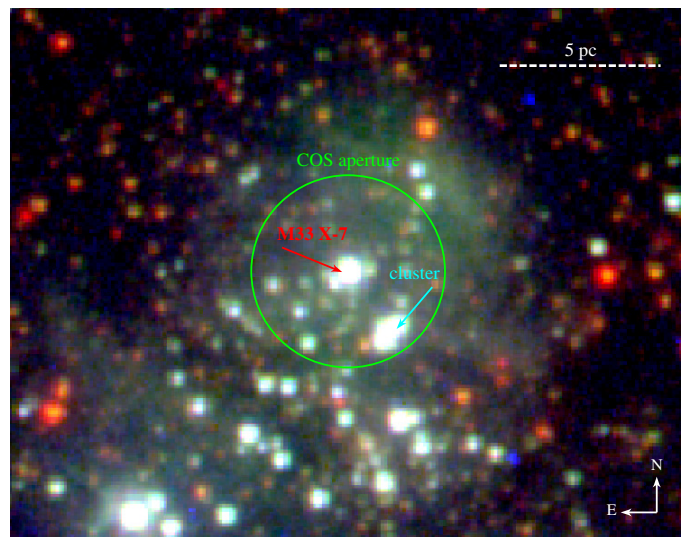}
\caption{\hst\ COS aperture around M33\,X-7. The background image is an RGB composite of \hst\ images in F336W, F475W, and F814W bands. A nearby cluster falls within the FOV of the COS spectrograph. }
\label{fig:aperture}
\end{figure}

\paragraph{\emph{Crowding in the COS aperture}:}

 Since M33\,X-7 resides in an OB association \citep{Ma2001,Pietsch2006}, separating the contamination of the COS UV spectrum by nearby sources is  challenging. The COS circular aperture is 1.25 arcsec in radius, but the aberrated beam entering the aperture allows objects up to 2 arcsec from the center to contribute to the recorded target spectrum. The most noticeable contribution comes from a bright neighboring cluster of stars, which is separated by only $0.9''$ to the southwest (see Fig.\,\ref{fig:aperture}). The position angle of the instrument during observation further reduced the distance along the cross-dispersion axis to $0.8''$. 
 
We extracted photometry of M33\,X-7 and the cluster from high-resolution \hst\ images in F275W, F336W, F475W, and F814W filters. The fluxes were measured using $0.2''$ aperture photometry and applied to the appropriate zero point. 
At 814\,nm, the brightness of the nearby cluster is just 0.2\,mag lower than M33,X-7, however at 275\,nm, the cluster becomes 0.6\,mag fainter than our target. In the FUV, only {\em GALEX} photometry of the region is available, with a spatial resolution of $\approx6''$. Even though M33\,X-7 dominates in the FUV, we expect a non-negligible contribution from the nearby cluster. This contamination could be larger in the optical and thus affect the parameters obtained by \cite{Orosz2007}.

In order to separate contributions from nearby sources in the COS aperture, we optimally extracted the one-dimensional target spectrum from the two-dimensional flat-fielded detector images. For this purpose we used observations carried out at $\phi=0.33$ and 0.53. Observation at these two phases were carried out with a position angle of $\approx116\degr$, implying a separation of $0.8''$ between our target and the nearby cluster along the cross-dispersion axis. At $\phi=0.95$ the position angle is $\approx96\degr$, so the corresponding separation is only $\lesssim0.6''$. 
In the two-dimensional spectra we could see a double peak profile on the cross-dispersion axis indicating contributions from both \mx\ and neighboring cluster. Hence we simultaneously fit for both our target and the cluster by multiplying them with the (wavelength-dependent) cross-dispersion instrumental profile and subsequently subtract from the combined observed spectra. This has been done along the cross-dispersion axis for all wavelengths. Finally, we extracted two spectral components such that the residual is minimum. We found that the disentangled spectra of M33\,X-7 have the same spectral features as the pipeline extracted spectra. Based on all exposures taken at $\phi=0.33$ and 0.53, we found that on average  M33\,X-7 accounts only for $\approx75\%$ of total flux, whereas the nearby cluster contributes the remaining 25\%.


\subsection{X-ray spectroscopy using \xmm}
\label{sec:xmm}

We carried out dedicated X-ray observations of M33\,X-7 with the \xmm\ telescope with a goal 
of obtaining X-ray spectroscopy  as close in time as possible to the {\em HST} observations 
in quadrature and inferior conjunction.
The log ofthe  \xmm\ observations is given in Table\,\ref{tab:observation}.

The \xmm\  data were analyzed using the Science Analysis System (SAS)
\footnote{\url{www.cosmos.esa.int/web/xmm-newton/what-is-sas}}. The standard analysis steps 
were performed to extract the spectra measured by the European Photon
Imaging Camera (EPIC). EPIC consists of a PN \citep{Struder2001} and two MOS \citep{Turner2001} CCD detectors in the focal plane, which operate in the
0.2--12 keV energy range. During our observations, all cameras were operating in the 
standard, full-frame mode with a thin filter. To analyze the EPIC spectra we used the standard 
spectral fitting software XSPEC \citep{Arnaud1996}.

\begin{figure}[t]
\centering
\includegraphics[width=1\columnwidth]{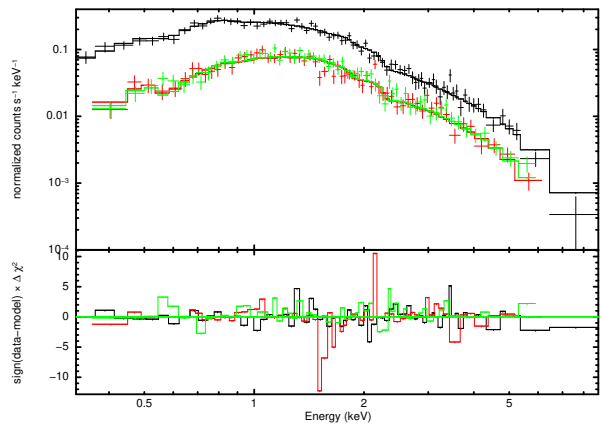}
\caption{\xmm\ EPIC PN,  MOS1, and MOS2
spectra of M33\,X-7  (black, red, and green curves, respectively)  obtained at phase $\phi\approx 0.47$. 
The error bars correspond to 3$\sigma$;  
the best-fit combined thermal plasma ({\it apec}) and {\it diskpbb} model is shown by 
solid lines. The model parameters are shown in Table\,\ref{tab:xmmspec}.  
The lower panel shows residuals between the data and the best-fit model.} 
\label{fig:201}
\end{figure}
\begin{figure}[t]
\centering
\includegraphics[width=1\columnwidth]{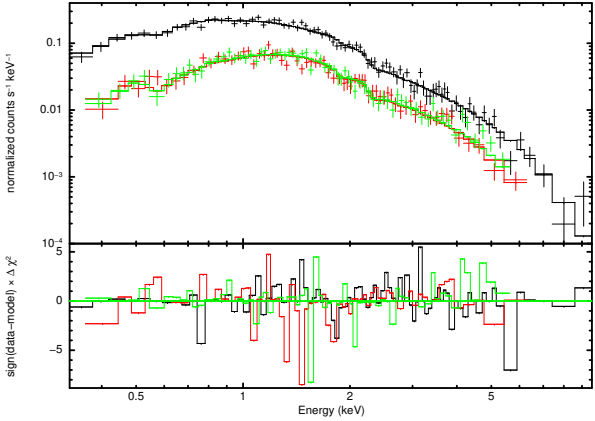}
\caption{Same as in Fig.\,\ref{fig:201}, but for phase $\phi\approx 0.33$.} 
\label{fig:401}
\end{figure}
\begin{table*}
\begin{center}
\caption[ ]{Parameters of the best-fit spectral models
to the observed  \xmm\ EPIC spectra of M33\,X-7. Two models are fitted. The first one is the plasma 
{\it apec} model for the metallicity $0.5\,Z_\odot$ combined  with a {\it diskpbb} model 
and corrected for ISM absorption {\it tbabs}.  The second one is the Bremsstrahlung 
({\it brems}) model corrected for ISM absorption {\it tbabs}.
Fluxes and luminosities are in the 0.2-12.0\,keV band. }
\begin{tabular}[]{lcc}
\hline
\hline
Model parameter & Best-fit values for $\phi\approx 0.47$ &  Best-fit values for $\phi\approx 0.33$ \\
\hline
\multicolumn{3}{c}{\it tbabs(apec+diskpbb)} \\
\hline
$N_{\mathrm H}$ [$10^{21}$ cm$^{-2}$]   & $2.3 \pm 0.3$ & $2.3\pm 0.3$ \\
$kT$ [keV]                              & $0.3\pm 0.1$  & $0.3\pm 0.1$ \\
$EM$ [$10^{59}$ cm$^{-3}$]              & $5.4\pm 3.8$  & $2.9 \pm 2.8$ \\
$T_{\rm in}$ [keV]                      & $1.17\pm 0.07$ & $1.25\pm 0.08$ \\
Parameter $p$: $T(r)\propto r^{-p}$     & $0.54 \pm 0.03$ & $0.53 \pm 0.02$\\
{\it norm}                              & $0.014  \pm 0.006$ & $0.007 \pm 0.003$\\                            
reduced $\chi^2$  & with $201$ d.o.f. $1.0$ & with $217$ d.o.f. $1.0$ \\
Model $F_{\rm X}$ [\flux] & $1.1\times 10^{-12}$ & $9.7\times 10^{-13}$ \\
Luminosity $L_{\rm X}$ [erg\,s$^{-1}$] & $1.7\times 10^{38} $ & $1.5\times 10^{37} $ \\
\hline
\multicolumn{3}{c}{\it tbabs*brems} \\
\hline
$N_{\mathrm H}$ [$10^{21}$ cm$^{-2}$]    & $2.1\pm 0.1$ & $1.9 \pm 0.1$   \\
$kT$ [keV]                               & $2.34\pm 0.08$ & $2.47\pm 0.09$  \\
$EM$ [$10^{60}$ cm$^{-3}$]               & $4.5\pm 0.1$ &  $3.6\pm 0.1$ \\
Reduced $\chi^2$  &  with $204$ d.o.f. $1.1$ &  with $220$ d.o.f. $1.05$ \\
Model $F_{\rm X}$ [\flux] & $1.1\times 10^{-12}$  & $9.9\times 10^{-13}$ \\
Luminosity $L_{\rm X}$ [erg\,s$^{-1}$] & $1.7\times 10^{38} $ & $1.5\times 10^{38} $ \\
\hline
\hline
\end{tabular}
\label{tab:xmmspec}
\end{center}
\end{table*}

The extracted spectra and the best fit spectral models are shown in Figs.\,\ref{fig:201} and \ref{fig:401}, while the best fit model parameters are listed in Table\,\ref{tab:xmmspec}. It is interesting to note that unlike HMXBs with neutron star companions accreting from the stellar wind of OB-type supergiants  \citep{Gimenez-Garcia2015,Sanjurjo-Ferrin2021}, 
no Fe\,K$\alpha$ line is seen in the \xmm\ EPIC spectra of M33\,X-7. 
After testing various models often used for describing the spectra of accreting BHs, the simplest model which well describes the spectra is a combination of thermal plasma {\em apec} and a multiple blackbody disk model {\em diskpbb}. The latter model is characterized by the inner disk temperature $T_{\rm in}$ and the scaling of the local disk temperature depends on the radius, 
$T\propto r^{-p}$, with $p$ being a free parameter \citep{Kubota2005}.

In the observation at phase $\phi\approx 0.47$, i.e.\ close to the inferior conjunction when the observer has the direct view on the accreting BH, the emission measure of the collisionally ionized plasma (described by the {\em apec} model) appears to be higher than at $\phi \approx 0.33$, i.e. close to quadrature. We interpret this {\em apec} plasma component as originating in the accretion disk wind. Furthermore, close to the inferior conjunction, the parameter {\em norm} which scales with the accretion disk inclination as $\cos\theta$  (with $\theta = 0$ being face-on) is twice as high as quadrature. This might imply that we see a larger fraction of the disk at inferior conjunction.  Interestingly, the X-ray absorption, as characterized by the parameter $N_{\rm H}$, does not change between the two orbital phases. This implies an insignificant contribution to the X-ray absorption from the general wind of the O-star beyond the BH orbit which could be due to the wind quenching by strong X-ray irradiation (see Sect.\,6.1)

\subsection{Archival optical spectra}
We made use of the optical spectra of M33\,X-7 published by \cite{Orosz2007} obtained from August 18 to November 16, 2006, with the 8.2-m Gemini North Telescope at wavelengths $\lambda$ = 4000-5000\,\AA. The final spectra is an average of spectra taken over 22 nights, and used these data for our spectral analysis. Although the optical data is not phase-resolved like UV or X-ray, it is sufficient for our spectral analysis since we do not expect any notable variation in the absorption lines at the blue optical range.


\section{Stellar atmosphere modeling}

We carefully analyzed the UV spectra obtained at three different phases (eclipse, quadrature, and inferior conjunction). As a first approximation, we used the stellar parameters obtained by \cite{Orosz2007}, where they used plane-parallel models from the TLUSTY grid to analyze the spectra. This approximation is only suitable for stars without winds where the spectrum is purely formed below the sonic point in a quasi-hydrostatic photosphere. However, since the donor in M33\,X-7 is an O giant or supergiant with a dense wind, the spectrum is altered in the layers beyond. This can cause an impact on the derived stellar parameters such as $\log\,g$. So-called unified model atmospheres are necessary to consistently describe the outermost layers of the star and their winds \citep[e.g.,][]{Gabler1989,Sander2017}.  Finally, X-rays need to be accounted for the spectral analysis of the HMXB donor since they can have a noticeable effect on the ionization structure of the wind. In turn this affects the derived parameters of the donor and its wind.

\begin{figure} 
	\centering
	\includegraphics[width=0.45\textwidth]{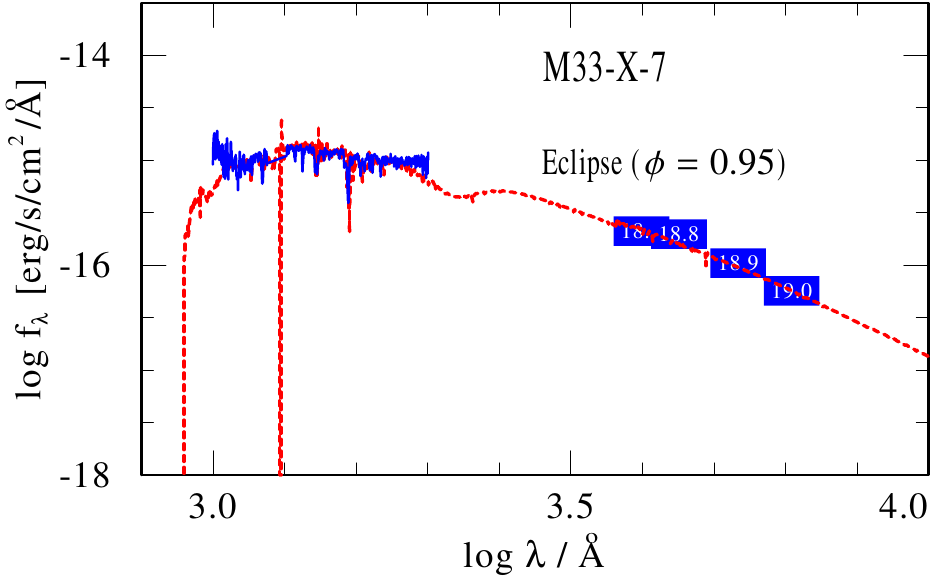} 
	\caption{Spectral energy distribution of M33\,X-7. The model SED (red solid
line) is adjusted to fit the photometry in UV and optical bands (blue boxes).}
	\label{fig:m33x7_sed}
\end{figure} 

\begin{figure}
	\centering
	\includegraphics[scale=0.75]{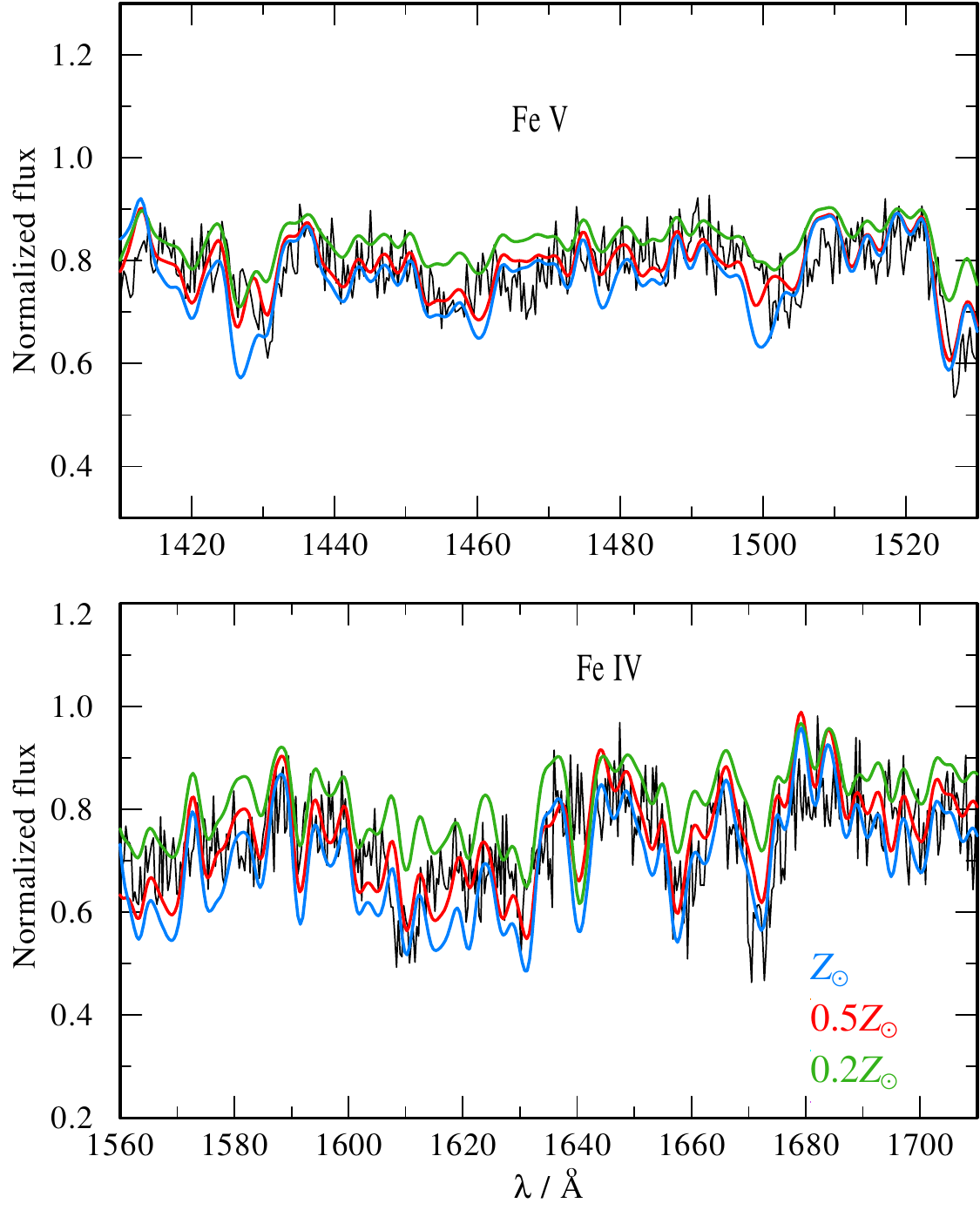}
	\caption{Iron forest in the FUV range observed in the M33\,X-7 spectra (black line). Over plotted in blue, red, and green are synthetic spectra obtained for models with the same parameters but with different iron abundances, ranging to 1, 1/2, and 1/5 Fe$_{\odot}$ respectively.}
		\label{fig:iron}
\end{figure}

To analyze the UV and optical spectra, we use the Potsdam Wolf–Rayet (PoWR) model\footnote{http://www.astro.physik.uni-potsdam.de/PoWR/}, which is a state-of-the-art non-local thermodynamic equilibrium (NLTE) code. The PoWR code assumes a spherically symmetric outflow and accounts for  iron-line blanketing, wind inhomogeneities, a consistent stratification in the quasi-hydrostatic part, and irradiation by X-rays, all of which are necessary to measure stellar and wind parameters.  To achieve a consistent solution, the equations of statistical equilibrium and radiative transfer are iteratively solved to yield the population numbers, while accounting for energy conservation. The radiative transfer is solved in the comoving frame, which avoids simplifications such as the Sobolev approximation. Once an atmosphere model is converged, the synthetic spectrum is calculated via a formal integration along emerging rays. Details of the PoWR code are described in \cite{Graefener2002},  \cite{Hamann2003}, \cite{Todt2015}, and \cite{Sander2015}. However, it should be noted that the standard version of the PoWR code does not account for deviations from spherical geometry.

A PoWR model is specified by the stellar temperature $T_\ast$, the bolometric luminosity $L$, the surface gravity $g_\ast$, the mass-loss rate $ \dot{M} $, the  wind terminal velocity $\varv_\infty$, the velocity law, and the chemical abundances.  The stellar radius $R_\ast$ (inner boundary) is defined at a Rosseland continuum optical depth of $\tau_{\mathrm {Ross}} = 20$ from $T_\ast$ and $L$. Following the Stefan-Boltzmann law $L = 4 \pi \sigma_{\mathrm{SB}}\, R_\ast^2\, T_\ast^4 $ the stellar temperature  $T_\ast$ is the effective temperature corresponding to the stellar radius $R_\ast$. The outer boundary in our models is set to $R_{\mathrm{max}} =100 R_\ast$.

In the main iteration, thermal broadening and turbulence are approximately accounted for by assuming Gaussian line profiles with a Doppler width of 40\,km\,s$ ^{-1} $. In the formal integral for the calculation of the emergent spectrum, the Doppler velocity is split into the depth-dependent thermal velocity and a microturbulence velocity $\xi(r)$. A microturbulent velocity of 10\,km\,s$ ^{-1} $ was adopted in the photosphere, growing proportional with the wind velocity.

The velocity field in PoWR models consists of two parts. In the inner part of the stellar atmosphere, the velocity field is calculated consistently such that the quasi-hydrostatic density stratification is fulfilled \citep{Sander2015}. In the supersonic region, the wind velocity field $\varv(r)$ is prescribed assuming a so-called $\beta$-law $\varv(r) = \varv_\infty \left( 1- R_\ast/r \right)^{\beta}$ \citep{CAK1975}, where $ \varv_\infty$ is the terminal wind velocity.  

In PoWR models, wind inhomogeneities are accounted for in the microclumping approach that assumes optically thin clumps \citep{Hillier1991,Hamann1998} with a void interclump medium and described by a so-called density contrast $D(r)$. The matter density in the clumps is enhanced by a factor $D = 1/f_\mathrm{V}$, where $f_\mathrm{V}$ is the fraction of volume filled by clumps. In the current study, we account for a depth-dependent clumping factor.

The models account for complex atomic data of H, He, C, N, O, Mg, Si, P, S, and the iron-group elements. Iron-group elements with millions of lines are included in the PoWR code through the superlevel approach \citep{Graefener2002}.


The PoWR code can account for ionization due to X-rays. The X-ray emission is modeled as described 
by \cite{Baum1992}, assuming that the only contribution to the X-ray flux is coming from free--free 
transitions. Since the current generation of PoWR models is limited to spherical symmetry,
the X-rays are assumed to arise from an optically-thin spherical shell around the star.
The X-ray emission is specified by three free parameters, which are the fiducial temperature 
of the X-ray emitting plasma $T_{\rm X}$, the onset radius of the X-ray emission $R_0$ $(R_0 > R_\ast)$, 
and a filling factor $X_{\rm fill}$, describing the ratio of shocked to nonshocked plasma.  
We set the onset radius to the orbital distance between the donor star and the BH.
To obtain the fiducial temperature which describes the spectral shape in the X-ray regime, 
we fit the X-ray spectra with the {\em brems} model. The resulting temperatures are 
shown in Table\,\ref{tab:xmmspec}. As the next step, the X-ray filling factor is adjusted such that 
the wavelength integrated X-ray flux from the observations are reproduced by the model. This is a similar procedure as adopted in \cite{Hainich2020}.

\section{Stellar parameters}

\begin{table}[!hbt]
	\caption{Stellar and wind parameters derived for M33\,X-7 in this work. The estimated or assumed mass fractions of elements are also listed. By adopting the derived parameters of the donor and orbital parameters from literature (see Table\,\ref{table:systemparameters}), we calculated further system parameters. See text for more details.}
	\label{table:parameters}
	\centering
	\renewcommand{\arraystretch}{1.4}
	\begin{tabular}{lc}
		\hline 
		\hline
		\vspace{0.1cm}
		Parameter                                         &          value       \\
		\hline
		\vspace{0.1cm}
		Spectral type                                  & O9 II                   \\ 
		$T_{\ast}$ (kK)                                & $31^{+2}_{-2}$          \\
		$T_{2/3}$\tablefootmark{\textdagger} (kK)                                 & $31^{+2}_{-2}$                    \\
		$\log g_\ast$ (cm\,s$^{-2}$)    & $3.4^{+0.2}_{-0.2}$     \\
		$\log L$ ($L_\odot$)                           & $5.54^{+0.1}_{-0.1}$    \\
		$\beta$                                        & $0.8^{+0.2}_{-0.2}$       \\
		$R_\ast$ ($R_\odot$)                           & $20.5^{+2}_{-2}$          \\
		$D$                                            & $40^{+10}_{-10}$         \\
		$\log \dot{M}$ ($M_\odot \mathrm{yr}^{-1}$)    & $-6.2^{+0.1}_{-0.2}$    \\
		$\varv_{\infty}$ (km\,s$^{-1}$)     & $1500^{+200}_{-100}$    \\
		$\varv \sin i$ (km\,s$^{-1}$)       & $250^{+30}_{-30}$       \\
		$\varv_{\mathrm{mac}}$ (km\,s$^{-1}$)       & $150^{+30}_{-30}$       \\
		$\xi$ (km\,s$^{-1}$)            & $10^{+5}_{-5}$          \\
		$X_{\rm H}$ (mass fr.)                         & $0.60^{+0.1}_{-0.1}$    \\
		$X_{\rm He}$ (mass fr.)                         & $0.395^{+0.1}_{-0.1}$    \\
		$X_{\rm C}/10^{-5}$ (mass fr.)                 & $8^{+2}_{-2}$           \\
		$X_{\rm N}/10^{-4}$ (mass fr.)                 & $5^{+2}_{-2}$         \\
		$X_{\rm O}/10^{-3}$ (mass fr.)                 & $2^{+1}_{-2}$           \\
		$X_{\rm Si}/10^{-4}$ (mass fr.)                & $2^{+1}_{-2.0}$         \\
		$X_{\rm Mg}/10^{-4}$ (mass fr.)                & $2$\tablefootmark{*}                     \\
		$X_{\rm Al}/10^{-6}$ (mass fr.)                & $7.6$\tablefootmark{*}                     \\
		$X_{\rm P}/10^{-6}$ (mass fr.)                & $2$\tablefootmark{*}                   \\
		$X_{\rm S}/10^{-5}$ (mass fr.)                & $4.4$\tablefootmark{*}                     \\
			$X_{\rm Ne}/10^{-4}$ (mass fr.)                & $1.8$\tablefootmark{*}                     \\
		$X_{\rm Fe}/10^{-4}$ (mass fr.)                & $7^{+1}_{-2.0}$                     \\
		$E_{B-V}$ (mag)                                & $0.22^{+0.03}_{-0.04}$ \\
		$R_V$                                          & $2.4$                   \\
		$M_\mathrm{spec}$ ($M_\odot$)                  & $38^{+22}_{-10}$        \\
		$M_\mathrm{BH}$ ($M_\odot$)                  & $11.4^{+3.3}_{-1.7}$        \\
		$d_{\mathrm{BH}}$ ($R_\ast$)                                 & $1.75^{+0.2}_{-0.1}$     \\
		 inclination $i$ [$\degr$]      &   $65$   \\
		$\varv_\mathrm{orb}$ at BH (km\,s$^{-1}$) & $510^{+100}_{-40}$      \\
		$\varv_\mathrm{wind}$ at BH (km\,s$^{-1}$)                            & $760^{+100}_{-100}$ \\
		$R_\mathrm{rl}\,(R_\ast)$ & $0.84^{+0.2}_{-0.1}$\\
		$\log\,Q_{\mathrm H}$ (s$^{-1}$)   & 49   \\
		\hline
	\end{tabular}
	\tablefoot{{\textdagger} $T_{2/3}$ is the effective temperature which refers to the radius where
the Rosseland mean optical depth in the continuum is 2/3\\
\tablefoottext{*}{corresponds to typical values for OB stars in the LMC based on \citet{Trundle2007}}
}
\end{table}

We primarily used UV spectra taken at the time of eclipse to measure accurate stellar parameters of the donor star since in this phase we see the part of the star which is largely undisturbed by the BH. We also re-analyzed the optical spectra from \cite{Orosz2007} \revise{(see Appendix Figs.\,\ref{fig:opt}, \ref{fig:opt2} and \ref{fig:opt3})}.  We started the model calculations with the parameters provided by \cite{Orosz2007} and adjusted them to get the best fit. The best-fitting model is selected after several iterations of visual inspection and systematic variation of the parameters. We simultaneously reproduce the normalized line spectrum and the spectral energy distribution (SED) of M33\,X-7 using our best-fit PoWR model, as shown in Figs.\,\ref{fig:m33x7_sed}, \ref{fig:m33x7_095}, and \ref{fig:opt}. Details on the wind parameters are described in Sect.\,\ref{Sect:wind}.

The observed spectra are best reproduced by a model with $T_{\ast}$ = 31 kK and $\log g_\ast$= 3.4, which is in line with the spectral class of a late O giant. To constrain the temperature, we used photospheric diagnostic lines such as \ion{C}{iii} and \ion{Si}{iii} in the UV  along with \hei and \heii lines in the optical spectra. Our estimated values of  $T_{\ast}$ and $\log g_\ast$ are lower by 4\,kK and 0.35\,dex, respectively, than previously suggested by \cite{Orosz2007}. How the variations in $T_{\ast}$ and  $\log g_\ast$ affect different line profiles is depicted in \revise{Figs.\,\ref{fig:opt}, \ref{fig:opt2} and \ref{fig:opt3})}. Hotter models would result in stronger\,\heii lines in the optical and weaker \ion{C}{iv} and \ion{Si}{iv} lines in the UV. We measured the surface gravity using the pressure-broadened wings of the Balmer lines. Since the ionization balance also reacts on $\log g_\ast$,  we readjusted the temperature accordingly. Overall, our best-fit model agrees well with the observed UV and optical lines. Subsequently, the emergent spectrum of the PoWR model was convolved with a rotational velocity of 250\,km\,s$^{-1}$ and macro-turbulent velocity of 150\,km\,s$^{-1}$ to reproduce the shape of the observed line profiles. Taking into account the updated inclination of the system, the star rotates with $\varv_\mathrm{rot}/\varv_\mathrm{critical}\approx0.55$.

The luminosity, color excess $E_\mathrm{B-V}$, and the extinction-law parameter $R_V$  were determined by fitting the model SED to the photometry and flux calibrated UV spectra (see Fig.\,\ref{fig:m33x7_sed}). 
We have used UBV broad-band magnitudes from \cite{Pietsch2006} and R magnitude from \cite{Massey2006}.
The extinction  $A_V= R_V \times E_\mathrm{B-V}$ derived by the final SED fit is in agreement with that of \cite{Orosz2007}. Subsequently, the model flux is scaled with the adopted distance modulus of 24.62\,mag \citep[see][for more details]{Orosz2007}. The foreground Galactic extinction is set to 0.052\,mag \citep[according to the NED extinction calculator,][]{Schlegel1998}, and the intrinsic extinction in M33 is negligible according to the detailed dust maps by \cite{Hippelein2003}. The derived luminosity was found to be lower than obtained by \cite{Orosz2007}.

\begin{figure}
	\centering
	\includegraphics[width=0.45 \textwidth]{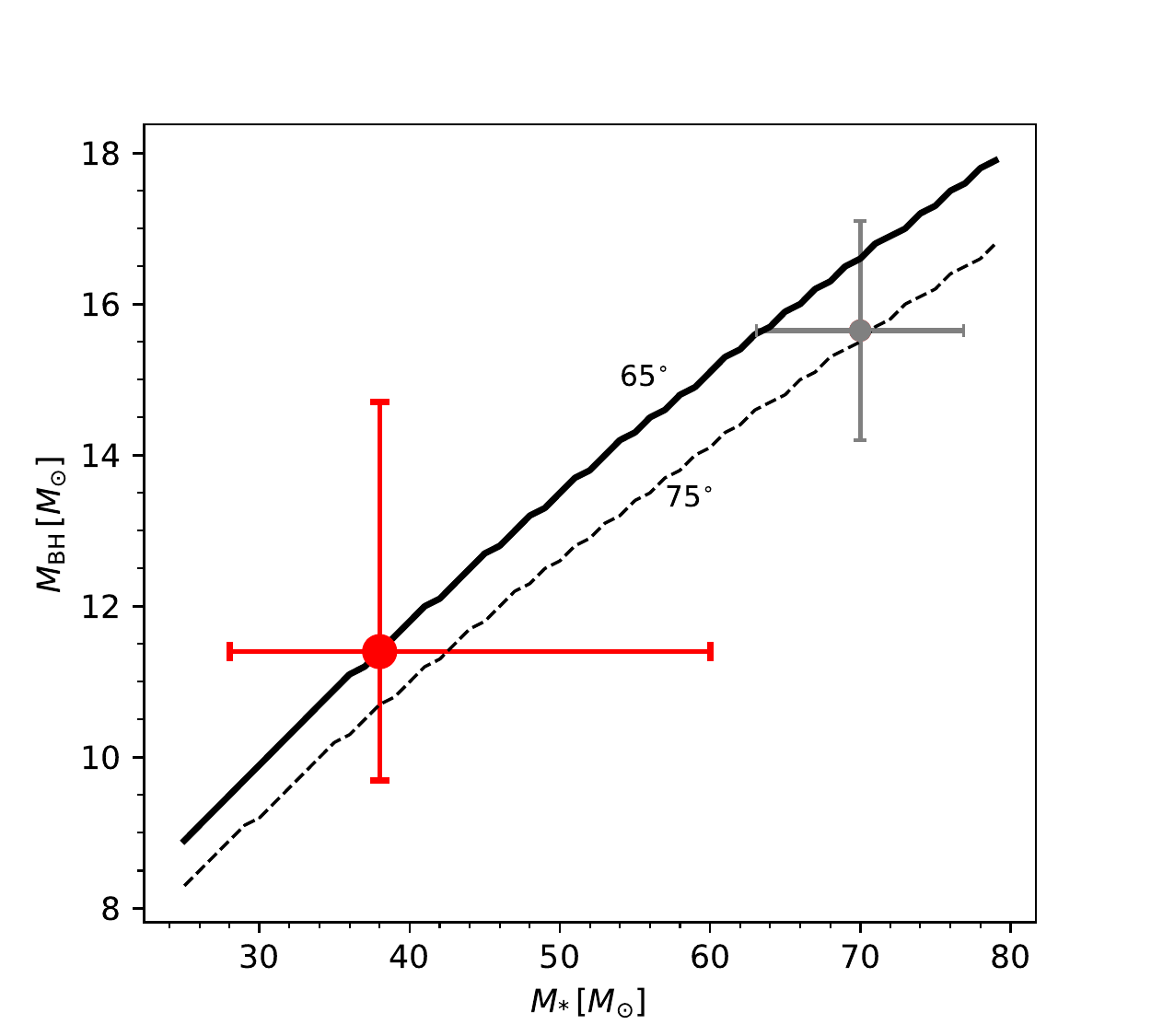}
	\caption{Mass plane diagram for components in the M33\,X-7. Masses derived in this work are marked in red whereas gray points represent previously determined values from \cite{Orosz2007}. The uncertainty in masses is shown as error bars. The solid lines represent loci of constant orbital inclination angle of $65\degr$ derived in this work. The dashed line corresponds to the inclination angle of $75\degr$ derived by \cite{Orosz2007}.  }
	\label{mass}
\end{figure}
\begin{figure*}
    \centering
    \includegraphics[width=\textwidth]{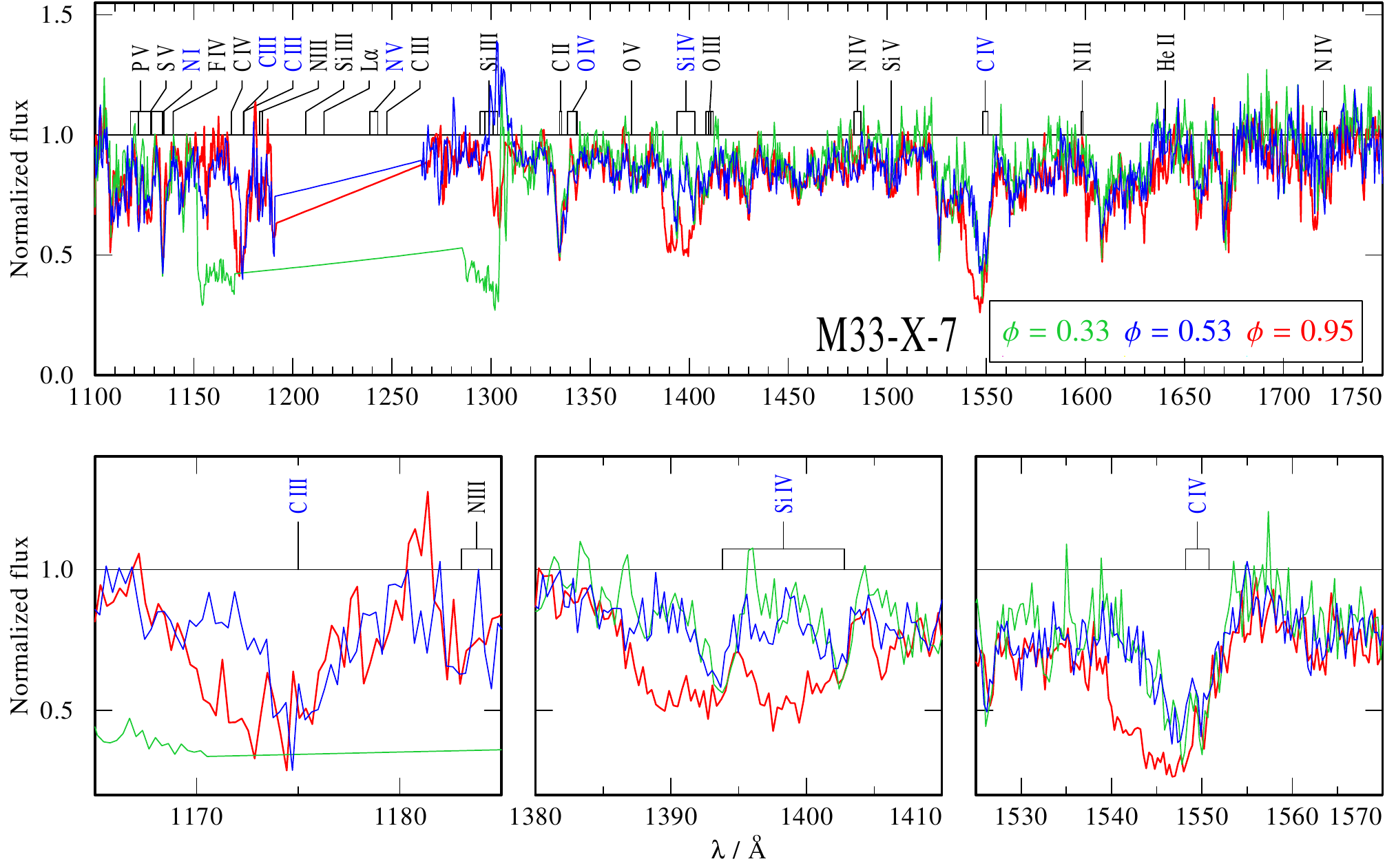} 
    \caption{Comparison of HST/COS spectra of M33\,X-7 taken during different orbital phases. Line colors indicate different orbital phases as given in the legend. The upper panel shows the full spectra range and lower panels focus on wind line profiles. The observed line variations are due to H-M effect (see Sect.\,\ref{Sect:HM}). }
    \label{fig:m33x7_phases}
\end{figure*} 

The Iron forest present in the FUV is most useful to constrain the metal content of the stars. Previous studies assumed a low metallicity of $Z=0.1\,Z_\odot$ for the system. However, we find that the iron forest in the UV range is best reproduced with models assuming an LMC-like metallicity of $Z=0.5\,Z_\odot$ (see Fig.\,\ref{fig:iron}). Considering the galactocentric distance of $\approx$ 2\,kpc from the nucleus of M33, this is in agreement with ISM abundances reported in the M33 star-forming regions \citep{Magrini2007,Alexeeva2022}.
 
We adjusted the abundances of individual elements to reproduce the observed strength of their respective absorption lines. The $\alpha$-element abundances are also in agreement with LMC-like metallicity.  We found evidence for nitrogen enrichment and carbon deficiency in the observed spectra. We needed to reduce the hydrogen mass fraction to 60\% and enrich helium up to 40\% to account for the strength of the \hei\ lines in the optical spectra. The He enrichment could be a result of the fast rotation of the donor. Alternatively, the outer layers might have been stripped by interaction with the BH companion. The UV spectra also show indications for CNO processing in the star. In order to account for the line strengths, we had to reduce the carbon abundance and increase the nitrogen abundance by a factor of five from baseline values. Individual element abundances and parameters derived from the spectral fitting are tabulated in Table.\,\ref{table:parameters}. For elements  Mg, P, S, Al, and Ne, we adopted typical LMC abundance values derived from OB stars \citep{Trundle2007}.

\paragraph{The masses of the components in M33\,X-7:} 
 The reduction in luminosity, temperature, and gravity resulted in a much lower spectroscopic mass of $\approx 38\,M_\odot$ for the O star than previously found in \cite{Orosz2007}. \revise{The spectroscopic mass is calculated from $\log\,g_\ast$ and $R_\ast$ ($g_\ast=G\,M_\ast\,R_\ast^{-2}$), and the uncertainties in these parameters  propagates into the final uncertainty of the donor mass.} The newly estimated mass falls in a similar range as that of O star donors in other BH HMXBs such as Cyg\,X-1, and LMC\,X-1 \citep{Miller-Jones2021Sci,Orosz2009}.  
 
\revise{Considering the orbital parameters derived by \cite{Orosz2007},  we estimated the BH mass, the orbital radius and the semi-eclipse angle for different values of the inclination angle. We found that an inclination angle of 65.5$\degr$ best matches the observed semi-eclipse angle of 26.5$\degr$ \citep{Pietsch2006}. If we take into account  the effects of an extended wind from the O-star, the donor radius slightly increases, therefore reduce the inclination angle to 65$\degr$. In comparison, the inclination angle derived in \cite{Orosz2007} was 74.6$\degr$.  The newly derived BH mass is $\approx 11\,M_\odot$, which is much lower than the previous estimate of $\approx 15.65\,M_\odot$. The scaling between the estimated masses of the donor and the BH along with their respective uncertainties are illustrated in Fig.\,\ref{mass}. }
 
 The mass ratio of the system is now $\approx3.3$. Based on the revised masses, we recalculated the orbital separation to $35.3\,R_\odot$, whereas the radius of the donor is $\approx20.5\,R_\odot$. The distance to the BH is always the same as the orbital separation, since the system has a nearly circular orbit. According to our results the BH is located much closer to the O giant than suggested by \cite{Orosz2007}.  Abundances and stellar parameters of the O star indicate that it is close to the main-sequence turn-off, overfilling its Roche-lobe radius.

\section{Orbital modulations in the UV wind lines}
\label{Sect:HM}

In HMXBs,  the X-ray source produces an extended zone of high ionization that moves around the orbit.  When this region is in front of the donor ($\phi$ = 0.5), the stellar wind gets over-ionized by X-ray irradiation. Therefore the intensity of the UV resonance lines is reduced due to the depletion of the absorbing ion species, while the opposite occurs at X-ray eclipse.  These orbital variations were first suggested by \cite{Hatchett-McCray1977} and they are expected to be observable in the ultraviolet resonance lines of ions such as \ion{C}{iv}, \ion{Si}{iv}, or \ion{N}{v}.

In Fig.\,\ref{fig:m33x7_phases} we present the UV spectra of M33\,X-7 taken at three different orbital phases. 
We identified  prominent P-Cygni profiles in the resonance lines of  \ion{C}{iii} at 1176\,\AA, \ion{Si}{iv}
$\lambda\lambda$1393-1403 and \ion{C}{iv} $\lambda\lambda$1548-1551 at X-ray eclipse. By comparing UV and optical spectral features to that of normal O stars, we re-classify the donor to an O9\,II spectral type.  We notice a significant change in these P-Cygni profiles at phases when the compact object is in the line of sight or in egress compared to that during eclipse. Such phase-dependent profiles variation of UV resonance lines across different orbital phases can be attributed to the X-ray ionization of the wind. 

The Hatchett-McCray effect (HM effect) is not observed for all HMXB systems since the orbital configuration influences the appearance of the wind line profiles. We report a strong HM effect in M33\,X-7 (compared to Galactic HMXBs), in both \ion{Si}{iv} and \ion{C}{iv} lines, indicating that the donor wind is significantly disturbed by the accreting BH. During the eclipse, the donor star wind profiles are much stronger and broader whereas the other two phases display a large reduction in the extent and depth of the blue-shifted absorption component.
Although variations at \ion{Si}{iv} and \ion{C}{iv} are most prominent at high velocities (-500 to -1500\,km\,s$^{-1}$), we found notable changes also at low velocities (-100 to -500\,km\,s$^{-1}$). The variation between phase 0.33 and phase 0.53 is mostly negligible, except at lower velocities, suggesting that the wind ionization extends all the way toward the photosphere of the O-star when the BH is in the line of sight, leaving only a ``shadow wind'' \citep[see e.g.,][]{Blondin1994} unaffected. 

The redward emission component of the P-Cygni lines is significantly reduced or absent in the eclipse spectra compared to that found in typical O giants. We find evidence for an increase in the emission component during phase 0.33 and phase 0.53. The \ion{He}{ii} at 1640\,\AA\, emission is also found to be stronger when the BH is in the line of sight. The reduction in the strength of the red emission component in the UV resonance lines is possibly due to the loss of ions by strong X-ray photoionization, because the red emission component forms in the opposite part of the star facing the BH. 

\section{Modeling the donor wind}
\label{Sect:wind}

Winds of massive stars are characterized mainly by their mass-loss rate $\dot{M}$, terminal wind velocity $\varv_\infty$, velocity law exponent $\beta$, and clumping factor $D$. In this work, we used UV spectra taken at $\phi=0.95$ to yield parameters of the O star wind and spectra taken at $\phi=0.53$ and 0.33 to understand the wind quenching by X-rays. 

The main diagnostic lines used to determine the terminal velocity are \ion{Si}{iv} $\lambda\lambda$1393-1403, and \ion{C}{iv} $\lambda\lambda$1548-1551. The maximum blue edge velocity in the unsaturated P-Cygni lines is $\varv_\mathrm{edge}\approx 1900$\,km\,s$^{-1}$. We find a narrow absorption feature (narrow absorption component, NAC) at a blueshifted velocity of $\approx1500$\,km\,s$^{-1}$. Based on this, we adopted the terminal velocity to be 1500$\pm100$\,km\,s$^{-1}$, which is also in agreement with $\varv_\infty \approx 0.8\, \varv_\mathrm{edge}$ \citep{Howarth1989,Prinja1990}. The ratio of terminal velocity to the escape velocity of the donor is found to be $\varv_\infty/\varv_{\mathrm{esc}}\approx2$, lower than for typical O stars
\citep{Lamers1995,Kudritzki2000}.

\begin{figure*}
\centering
\includegraphics[width=\textwidth]{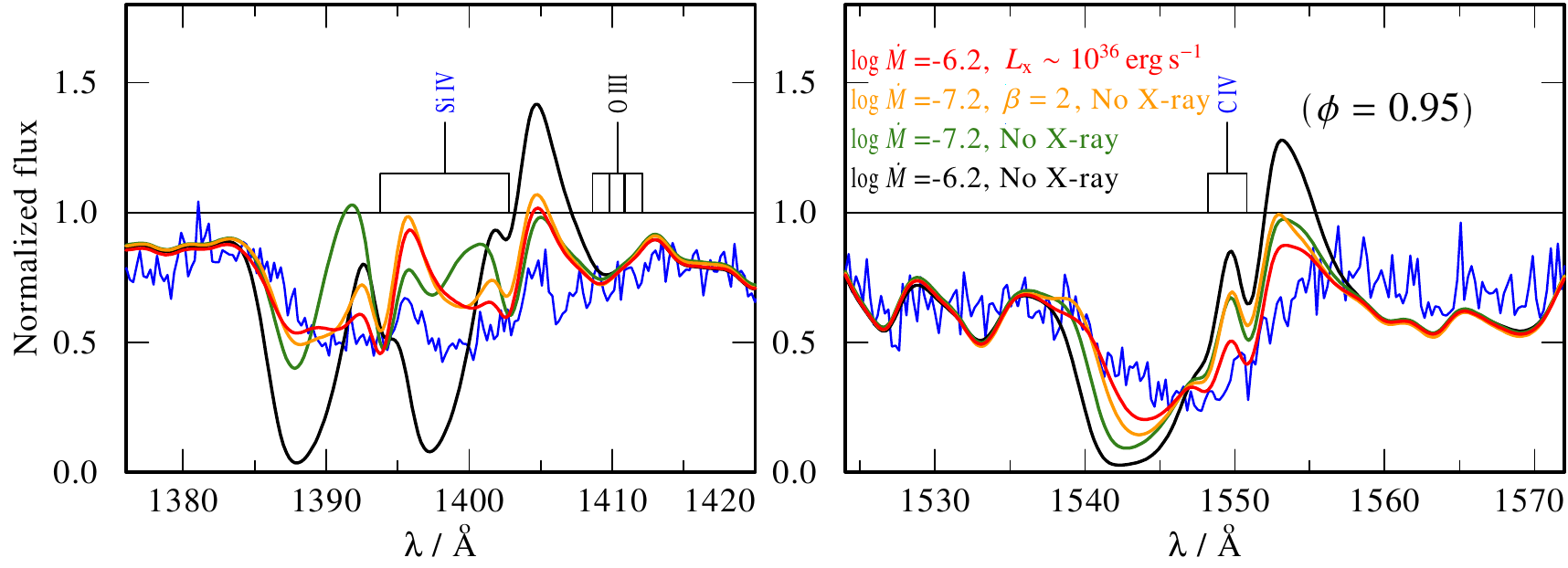} 
\caption{Comparison of observed \ion{Si}{iv} (left panel) and \ion{C}{iv} (right panel) line profiles (blue solid) at eclipse to the models calculated with different wind parameters and X-rays. Various colors represent changes in the adopted model parameters as given in the legend.  Please see text for more details.}
\label{fig:pcygni_095}
\end{figure*} 

\begin{figure*}
\centering
\includegraphics[width=\textwidth]{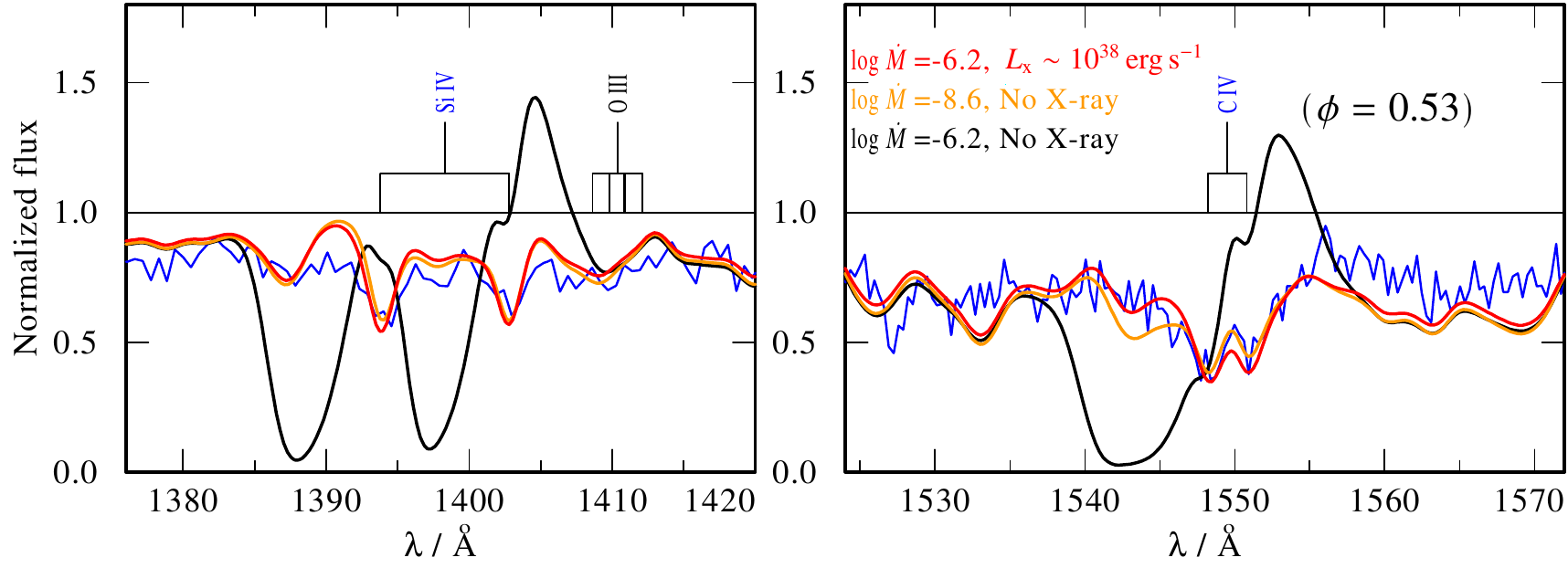} 
\caption{Same as Fig.\,\ref{fig:pcygni_095}, but at inferior conjunction}
\label{fig:pcygni_053}
\end{figure*} 
A microturbulent velocity of 10\,km\,s$^{-1}$ in the photosphere is derived based on the optical lines, and is also in agreement with the UV spectra. The turbulent velocity grows proportional to the terminal velocity in the wind regime. In the outer part of the wind, it reaches typically up to a value of  $\xi(R_\mathrm{max})=0.1\varv_\infty$. However, for M33\,X-7 we measure a higher turbulence velocity,
with a ratio $\xi(R_\mathrm{max})/\varv_\infty \approx$ 0.3.  The wind turbulent velocities derived here are larger compared to the previous studies of OB stars in the LMC and the Galaxy (e.g., 0.1 $\varv_\infty$ by \citealt{Kudritzki2000}, 0.14 $\varv_\infty$ by \citealt{Herrero2001}). A UV spectroscopy study of early B supergiants in M33 by \cite{Urbaneja2002} also reported high turbulent velocities in the range of 0.25-0.35 $\varv_\infty$.

 The adopted microclumping is depth-dependent, assumming that clumping begins at the sonic point, and increases outward. The maximum value of the clumping factor is found to be $D=40$, which is required to reproduce the UV lines, especially \ion{P}{v} $\lambda\lambda$1118-1128 and \ion{N}{iv} at 1718\,\AA. Our analysis suggested that the clumping factor reaches the maximum value only beyond 3$R_\ast$. Compared to this distance, the location of the BH is much closer (1.75$R_\ast$) to the photosphere of the donor star. Near the location of the BH, the wind clumping is lower ($D=10$).

One noticeable feature in the UV spectra are the weak unsaturated P-Cygni lines with negligible or no emission component. Whereas, OB supergiants with similar spectral types typically show \ion{Si}{iv} and \ion{C}{iv} resonance lines with saturated blue absorption and strong emission in the red part of the line.  We tried different approaches such as adjusting the mass-loss rate, the velocity law, the outer boundary of the wind, and including X-rays, to avoid the red emission part and to improve the fit quality to the observed UV lines. For that, we computed models with $\log \dot{M}$ in the range of -6.0 to -9.0. To model the supersonic part of the wind, we tested $\beta$ values in the range 0.6 to 3.0. A comparison of different possible solutions that can reproduce the observed spectra and our final best-fit parameters are discussed below.

\begin{figure*}
\centering
\includegraphics[width=\textwidth]{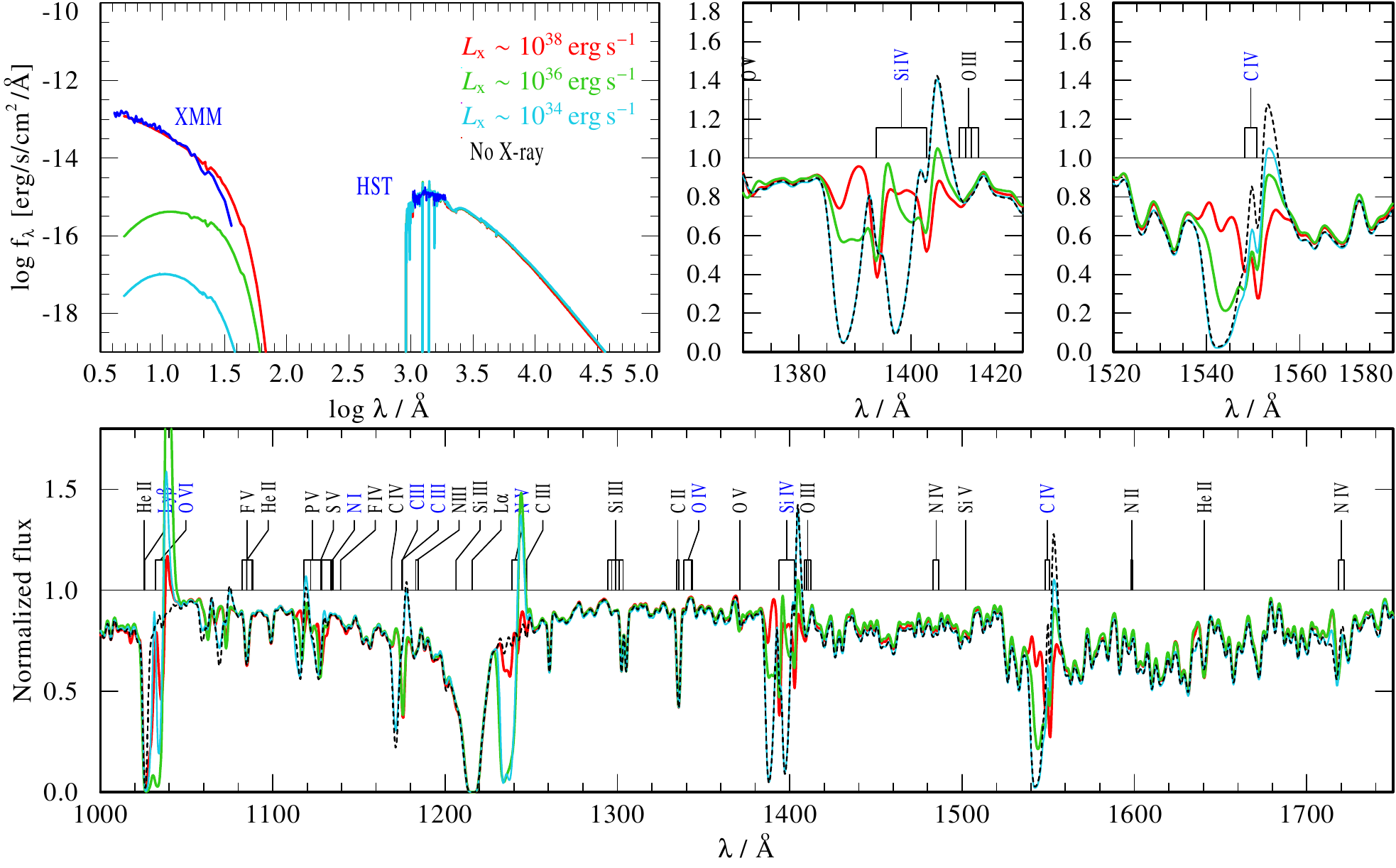}
\caption{Comparison of model SEDs (left upper panel) and emerging UV spectra (lower panel) for different X-ray illuminations. The model SEDs are compared with \xmm\ and \hst\ spectra taken at phase $\phi=0.53$. A zoom on \ion{C}{iv} and \ion{Si}{iv} line profiles are displayed in the upper panels (middle and right).  }
\label{fig:xmodel}
\end{figure*} 

\begin{figure}
	\centering
	\includegraphics[width=0.48 \textwidth]{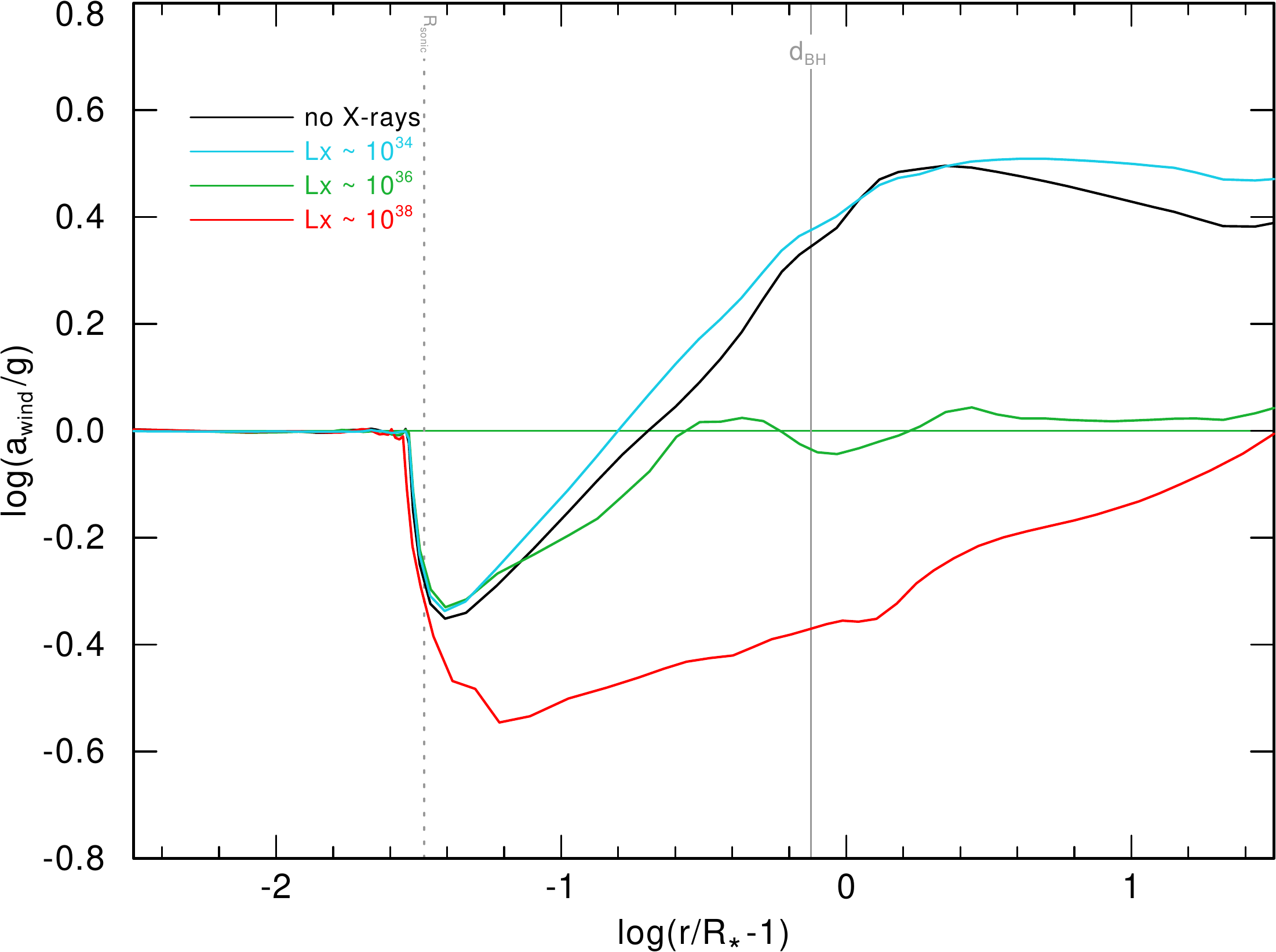}
	\caption{Comparison of total wind acceleration normalized to the gravitational acceleration for models including different amounts of X-rays. The location of the sonic and the BH radius are marked with dashed and solid gray lines.}
	\label{fig:acc_comp}
\end{figure}

As a first attempt, we tried to change the wind parameters in the model without incorporating X-rays. A model with a low mass-loss rate ($\log \dot{M}\approx-7.2$) assuming a shallow wind velocity law ($\beta$=2) is found to be in good agreement with the observations at phase 0.95 (orange line in Fig.\,\ref{fig:pcygni_095}). The observed profile can also be reproduced by models with a moderate mass-loss rate ($\approx$-6.6) and adopting a velocity law exponent of $\beta$=0.8 if we assume the wind to be shut off by X-rays beyond 900\,km\,s$^{-1}$. On the other hand, the theoretically predicted mass-loss rate from \cite{Vink2000,Vink2001} for the derived donor parameters is much higher, $\log \dot{M}\approx-6.2$. Assuming such a higher $\dot{M}$ and $\beta$=0.8 in the model would result in much stronger P-Cygni profiles compared to the \hst\ spectra taken during the X-ray eclipse (see the black line in Fig.\,\ref{fig:pcygni_095}). Models with a lower mass-loss rate would make the \ion{Si}{iv} P-Cygni lines much weaker, but not have much impact on the \ion{C}{iv} profile.

Compared to eclipse, the observed spectra taken at $\phi=0.53$ and 0.33 do not show any wind line features. Both \ion{Si}{iv} and \ion{C}{iv} appear to be mostly photospheric (see Fig.\,\ref{fig:pcygni_053}). To mimic the observed spectral features at this phase without accounting for the effect of X-ray photoionization, we need to adopt a very low mass-loss rate of $\log \dot{M}\lesssim-8.6$.



\begin{figure*}
	\centering
	\includegraphics[width=0.8 \textwidth]{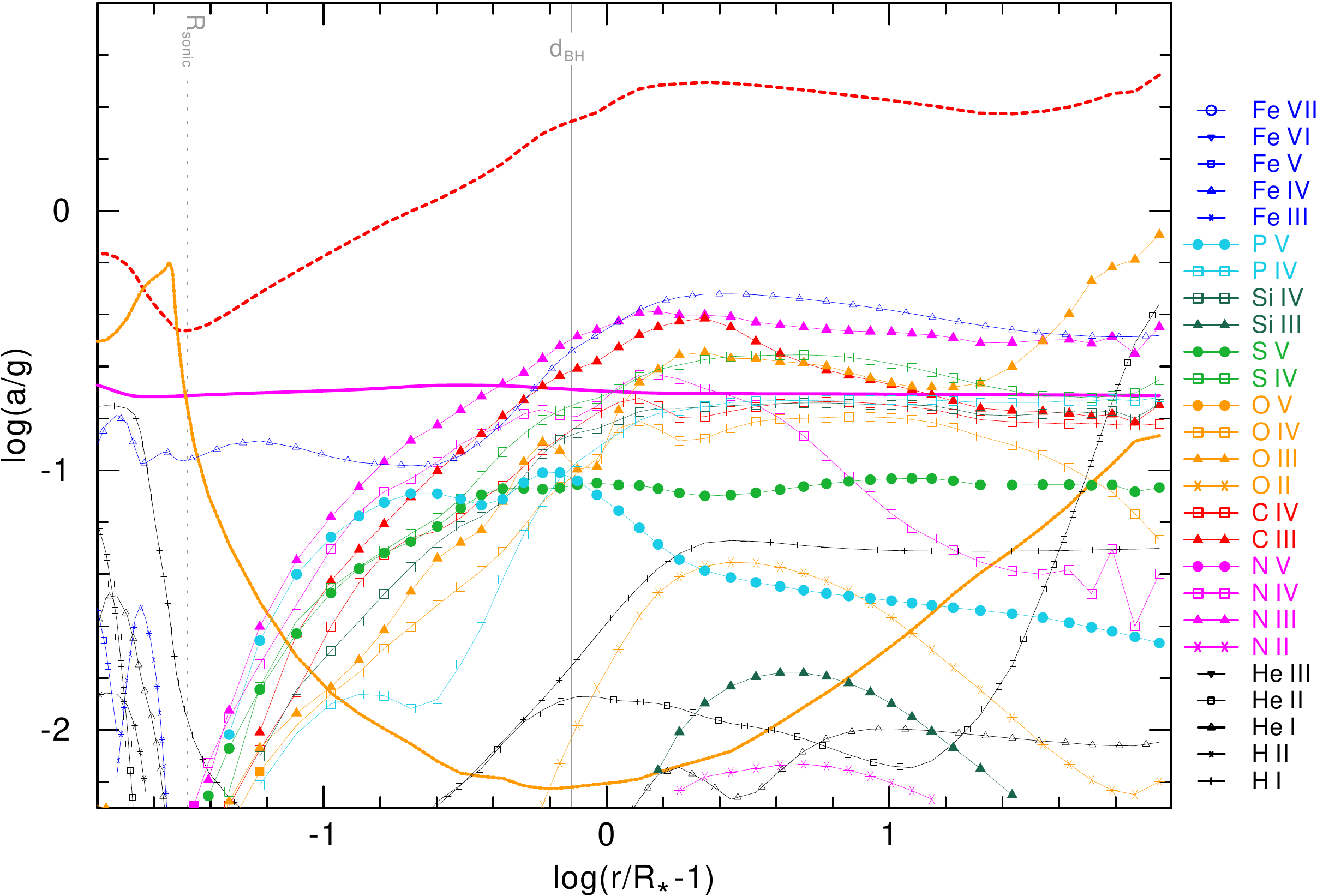}
	\includegraphics[width=0.8 \textwidth]{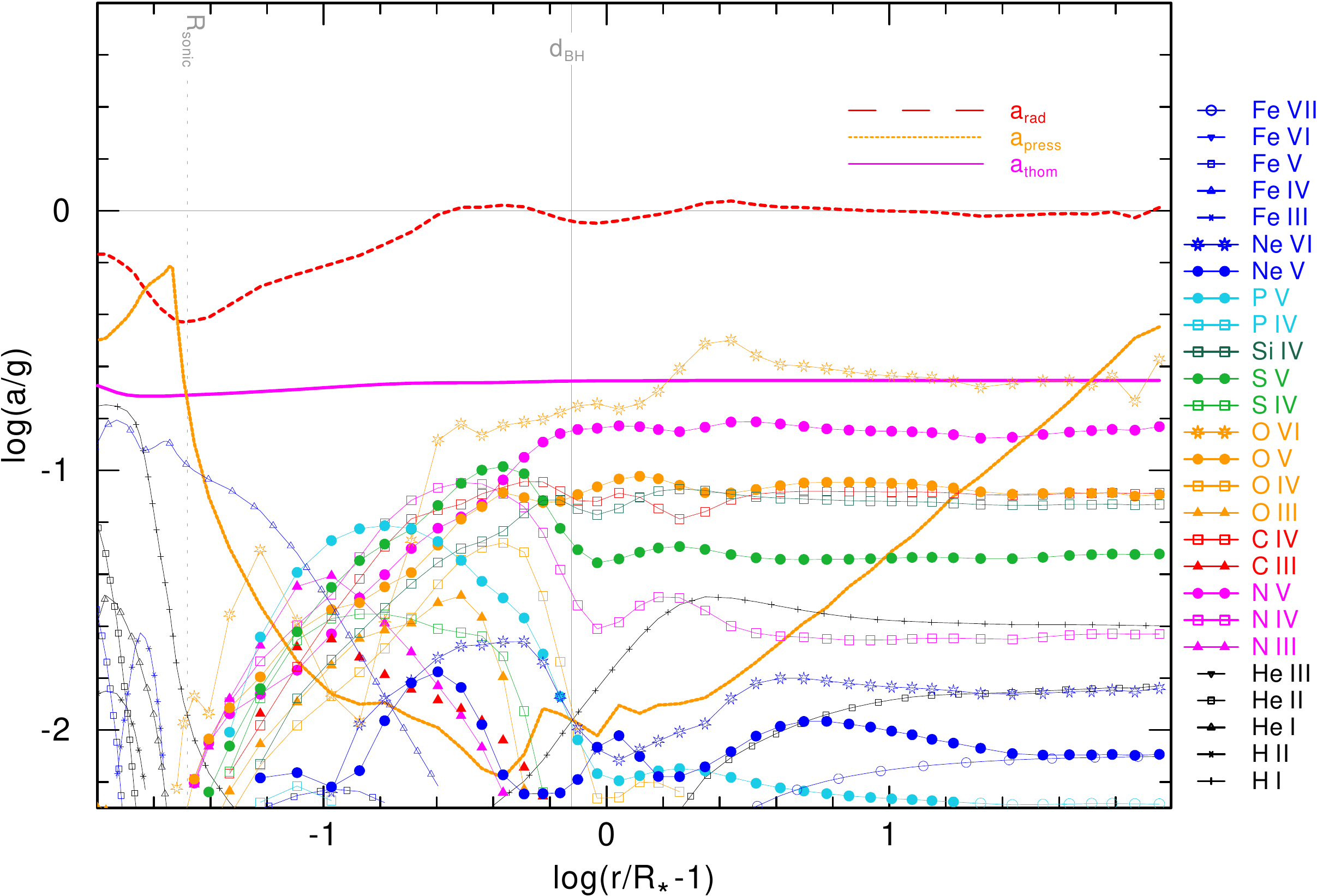}
	\caption{Contributions to the radiative acceleration from the different ions for models with no X-ray (top) and including $L_{\mathrm{x}}\sim 10^{36}$\,erg\,s$^{-1}$ (bottom). Symbols and colors representing different elements and their ionization state is shown in the figure. The plot is limited to the ions that contribute at least $\sim1$\% to the radiative acceleration. Line acceleration due to scattering by free electrons (magenta solid curve) and gas pressure (orange dashed curve) and total radiative acceleration (red dashed curve) are plotted for comparison. The radius at which the BH is located is marked by a gray solid line and the location of the sonic point marked by gray dotted line.}
	\label{ion1}
\end{figure*}
\begin{figure*}
	\centering 
	\includegraphics[width=0.8 \textwidth]{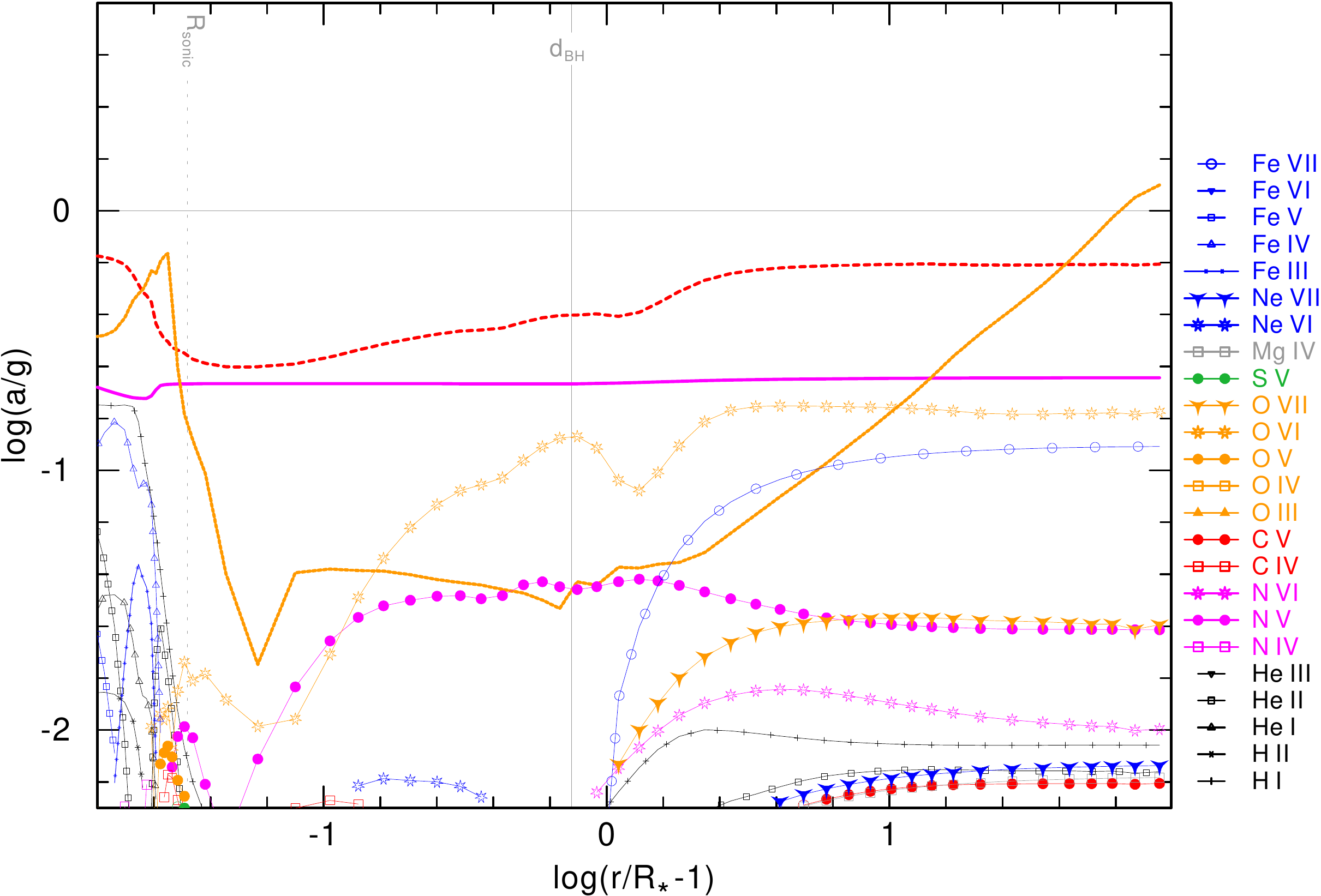}
	\caption{Same as Fig.\,\ref{ion1}, but for model including $L_{\mathrm{x}}\sim 10^{38}$\,erg\,s$^{-1}$ }
		\label{ion2}
\end{figure*}
\subsection{Impact of X-rays on the UV spectra}
\label{subsect:xray}
X-ray irradiation from the BH accretion disk makes a significant impact on the ionization stratification of the donor wind. While the photospheric absorption lines are unaffected,  the appearance of spectral lines that originate in the wind,  most notably in the UV range, changes significantly. Even adding a small amount of X-rays ($L_{\mathrm{x}}\lesssim 10^{32}$\,erg\,s$^{-1}$) in the model to account for wind intrinsic X-ray emission of the star, changes the ionization structure of the donor \citep[e.g.,][]{Oskinova2011}. However, the line profiles can get either stronger or weaker depending on the wind density near the BH and the X-ray intensity. To understand the impact of X-ray photoionization on the observed UV spectra, we have calculated models with different X-ray luminosities ($10^{32-38}$\,erg\,s$^{-1}$). In our spherically symmetric models, the X-rays are assumed to be generated from an optically-thin spherical shell around the star, whereas the geometry of the HMXB is more complex. In this system, the BH is behind the donor during the eclipse and in the foreground during other phases. On top of this, the radius of the donor star is large enough to fill its Roche-lobe (see also Sect.\,\ref{Sect:accretion}). Nevertheless, this is one of the first attempts to study the detailed wind driving of a massive star in a BH HMXB.

The effects of the X-ray field on the emergent spectra are illustrated in Fig.\,\ref{fig:xmodel}. The mass-loss rate is fixed to $\log \dot{M}\approx -6.2$ for all cases. The synthetic spectra generated by models with no X-rays and  moderate  X-ray irradiation ($L_{\mathrm{x}}\lesssim 10^{34}$\,erg\,s$^{-1}$) show strong saturated \ion{Si}{iv} and \ion{C}{iv} P-Cygni profiles.  Including X-rays makes a pronounced effect on wind lines of highly ionized species like \ion{O}{vi} and \ion{N}{v}. These lines are absent in models which do not account for X-rays, but the inclusion of X-rays causes sufficient photo- and Auger ionization to populate the \ion{O}{vi} and \ion{N}{v} states. 
The overall wind acceleration remains about the same for both cases (see black and cyan lines in Fig.\,\ref{fig:acc_comp}).
A further increase in the X-ray field to $L_{\mathrm{x}}\sim 10^{36}$\,erg\,s$^{-1}$  affects the emerging UV spectrum drastically. The strength of the wind lines (such as \ion{Si}{iv}, \ion{C}{iv}, \ion{C}{iii}, and \ion{P}{v}) reduces significantly whereas \ion{O}{vi} becomes stronger. A model including a strong X-ray field on the order of  $L_{\mathrm{x}}\sim 10^{38}$\,erg\,s$^{-1}$ considerably reduces the strength of \ion{Si}{iv} and \ion{C}{iv} lines, making them appear to be photospheric lines.

Since the important wind-driving ions are depopulated, the line acceleration in the outer wind decreases correspondingly. As shown in Fig.\,\ref{fig:acc_comp}, strong X-ray illumination causes a strong decline in the wind accelerations from the sonic radius onward and hence stalling of the wind. This is because photoionization destroys the ions that are capable of absorbing photons in the UV. The normalized wind acceleration drops to unity for models with  $L_{\mathrm{x}}\sim 10^{36}$\,erg\,s$^{-1}$.  The wind driving drops further below unity for models irradiated by  $L_{\mathrm{x}}\sim 10^{38}$\,erg\,s$^{-1}$ (see red line in Fig.\,\ref{fig:acc_comp}). 
The total wind acceleration is a factor of seven lower at the side of the donor facing BH and a factor of three lower in the shadow region compared to that of an unperturbed O-star. 
Our investigations thus suggest a breakdown of wind toward the BH in M33 X-7.
For this model, the blue \ion{O}{vi} and \ion{N}{v} absorption components only reach up to 1000\,km\,s$^{-1}$ instead of 1500\,km\,s$^{-1}$ measured at eclipse. The decrease of the wind terminal velocity due to X-ray irradiation is supported by previous HMXB studies \citep[see e.g.,][]{Watanabe2006,Sander2018,Krticka2018}.


\subsection{Wind driving}
\label{sec:winddriving}

In order to have a detailed picture of wind driving in these situations, we need to look at the contributions of the various elements in different ionization stages. This is illustrated in Figs.\,\ref{ion1} and \,\ref{ion2} by showing the radiative acceleration contribution from different ions and how this is affected by X-rays.
While these are not dynamically consistent models, the changes due to the X-ray illumination are representative nonetheless.
For a model without including X-rays (Fig.\,\ref{ion1} upper panel), the leading ions are \ion{Fe}{iv}, \ion{N}{iii}, and \ion{C}{iii}. \ion{O}{iii} becomes dominant in the far outer part of the wind. 
Near the sonic point, iron group opacities are the leading contributors to the wind along with the electron scattering and gas pressure. Beyond $R>1.1R_\ast$, CNO and other higher elements become important agents in wind driving. Higher ions such as \ion{Si}{iv},  \ion{C}{iv}, \ion{S}{iv} and \ion{N}{iv} also make a significant contribution to the radiative acceleration.  The contribution by C, N, O, and iron lines outweigh the free electron (Thomson) scattering contribution in the outer wind.  

In contrast, the model with  $L_{\mathrm{x}} \sim 10^{36}$\,erg\,s$^{-1}$ (Fig.\,\ref{ion1} lower panel) depopulates all the lower ions, especially beyond the radius at which the BH is located. In the inner part, the picture remains the same as for the model with no X-rays. Beyond $R>1.1R_\ast$, the acceleration by \ion{Fe}{iv} reduces rapidly and becomes unimportant. In this case \ion{O}{vi} and \ion{N}{v} are the leading contributors in the outer wind as we see can from the emerging UV spectra in Fig.\,\ref{fig:xmodel}. The total radiative acceleration by these ions is comparable to that of free electrons.  \ion{Si}{iv} and \ion{C}{iv} also make a noticeable contribution to the wind driving, but their influence is a factor of three lower compared to the model with no X-ray ionization. Increasing the X-ray field to $L_{\mathrm{x}}\sim 10^{38}$\,erg\,s$^{-1}$ makes a remarkable further difference as shown in Fig.\,\ref{ion2}. Strong X-ray photoionization results in the super-ionization of lower species hence only a few highly ionized species are contributing to the wind driving. \ion{O}{vi} is still the leading ion in this case, and other higher ions such as \ion{Fe}{vii}, \ion{N}{iv} and \ion{O}{vii} are also important. However, the contributions from these ions are lower than that by electron scattering. In comparison to the other two scenarios, here we can see a completely different picture in the inner part of the wind, where a plethora of ions contributing to the wind driving including the iron group components, are missing.
It should be noted that apart from CNO and Fe, elements like P, S, Si, Mg, and Ne also make important contributions in wind driving. Excluding higher elements and higher ionization stages would significantly affect the wind structure and the emerging spectra.

\begin{figure*} 
	\centering
	\includegraphics[width=\textwidth,trim={0 0cm 0 5.85cm},clip]{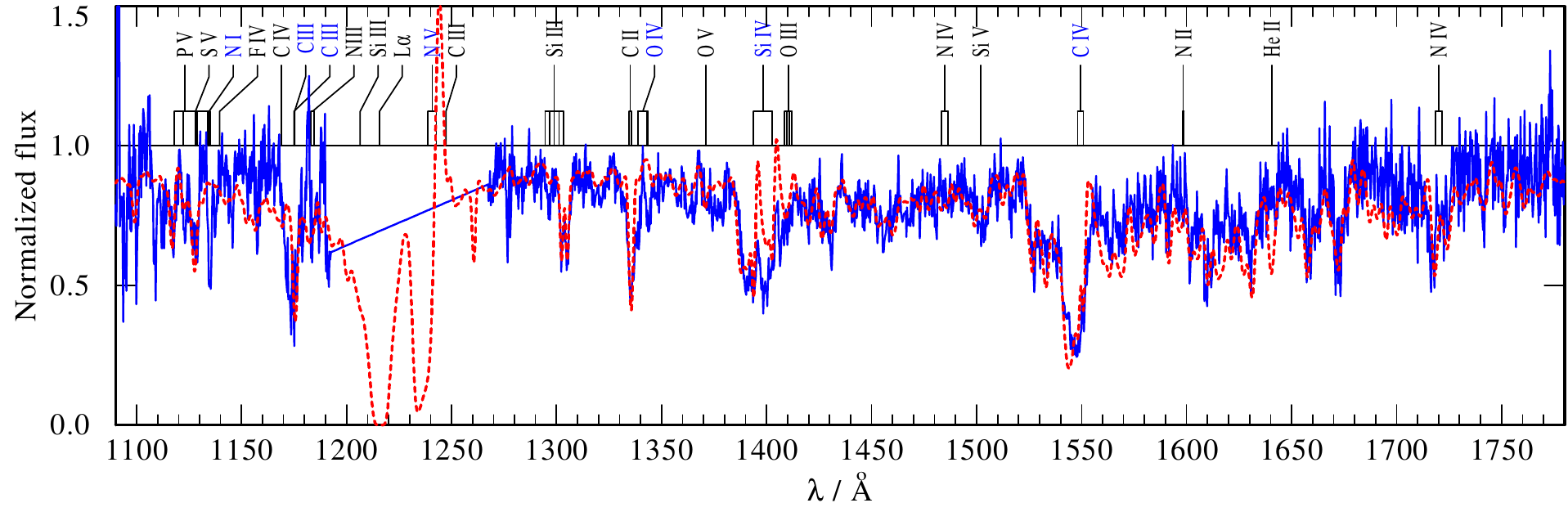} 
	\caption{Spectral fit for M33\,X-7 UV spectra at the X-ray eclipse.  Normalized \hst\ spectra (blue solid line) is overplotted with the PoWR model (red dashed line). The atmospheric parameters and abundances of this best-fit model are given in Tables\,\ref{table:parameters}.}
	\label{fig:m33x7_095}
\end{figure*} 

\begin{figure*}
    \centering
    \includegraphics[width=\textwidth,trim={0 0cm 0 5.85cm},clip]{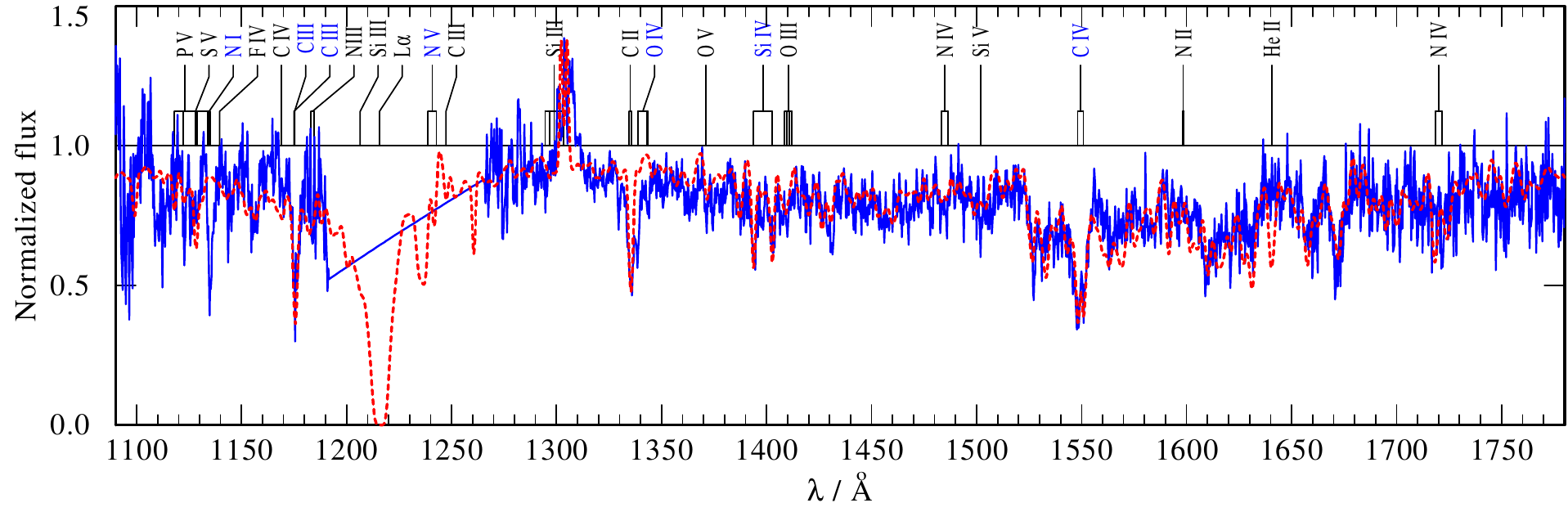} 
    \caption{Same as Fig\,\ref{fig:m33x7_095}, but spectra taken at inferior conjunction ($\phi\approx0.53$) where BH is in front of the donor star.  }
    \label{fig:m33x7_053}
\end{figure*} 

\begin{figure*}
    \centering
    \includegraphics[width=\textwidth,trim={0 0cm 0 5.85cm},clip]{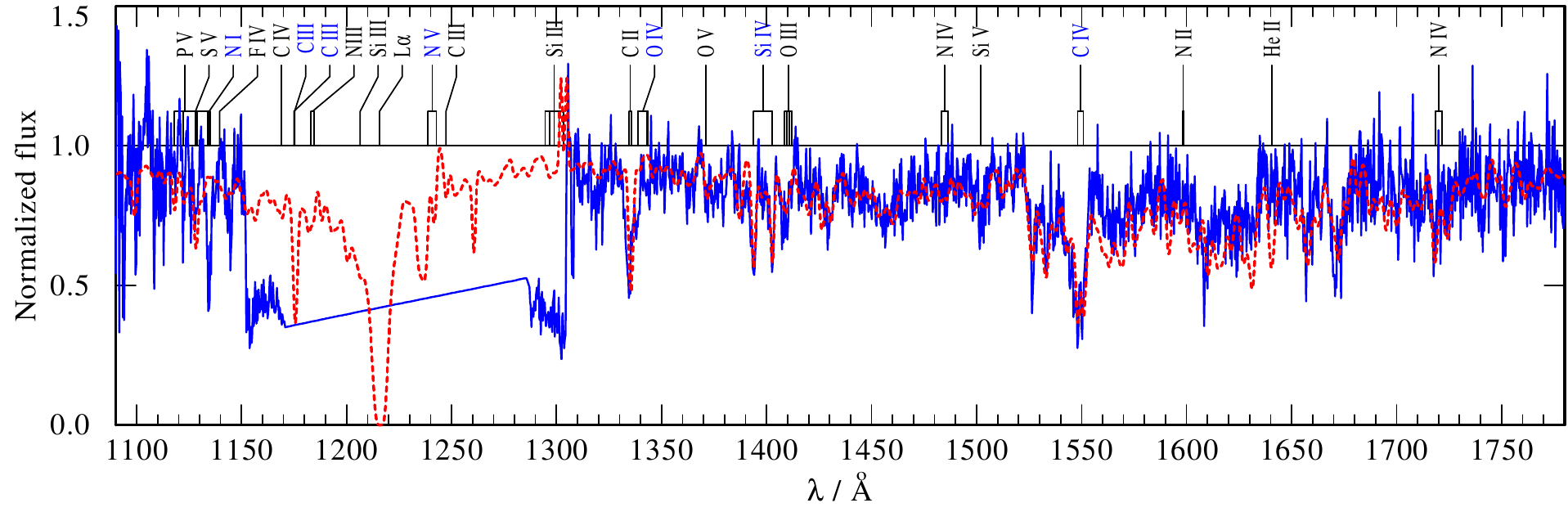} 
    \caption{Same as Fig\,\ref{fig:m33x7_095}, but spectra taken at Quadrature ($\phi\approx0.33$).  }
    \label{fig:m33x7_033}
\end{figure*} 

As the calculation of hydrodynamically consistent atmosphere models is beyond the scope of the present paper, one has to be careful with drawing deeper conclusions about the absolute radiative acceleration in the deeper wind layers. However, our studies give us very valuable insights regarding relative changes of the wind acceleration between the different situations. Without X-rays, the acceleration in the outer wind is approximately constant with $1 < \Gamma_\text{rad} \approx 3$, which is quite similar to the result for the much cooler Vela X-1 donor obtained with a dynamically consistent model \citep{Sander2018}. With the amount of X-rays measured during eclipse in M33 X-7, the situation reduces to $\Gamma_\text{rad} \approx 1$. Taking this result literally would imply that there is hardly any wind acceleration, but as we do see wind signatures in the UV spectrum during eclipse we know that $\Gamma_\text{rad} > 1$ must be fulfilled at least in an inner region where the wind would be launched and which we do not model dynamically consistent in this work. The consistent
model from \citet[their figure 6]{Sander2018} reflects such a situation where even for an illumination of $L_\text{X} \approx 10^{37}$
there is still a meaningful radiative acceleration close to the star before a complete breakdown occurs further out. Moreover, the inner acceleration in our M33 X-7 models is not affected when comparing the model without X-rays to those using the $L_\text{X}$ from the eclipse. This changes drastically, when looking at the strong irradiation case where the full $L_\text{X} \approx 10^{38}$ from the BH are irradiated. Now, even the acceleration in the inner region decreases by about 40\%. Close to the BH, the acceleration
further decreases by another 60\% compared to the eclipse situation. With $\Gamma_\text{rad} < 1$ now, the acceleration of the wind breaks down. For an isolated star, this would lead to a significant deceleration of the wind, as also discussed by \citet{Krticka2018,Krticka2022} for a wider range of system parameters. On top, the presence of the BH and its gravitational potential will need to be considered here,
which we will discuss in Sect.\,\ref{Sect:accretion}.

\subsection{X-ray luminous HMXB}
\label{sec:xlumhmxb}

The typical wind-intrinsic X-ray emission of an O-giant donor is $L_{\mathrm{x}}\sim 10^{32}$ \citep{Nebot2018}. However, this is much lower than the observed X-ray luminosity at eclipse estimated by \cite{Pietsch2004} using \xmm\ observations. They derived  $L_{\mathrm{x}}= 2-3 \times 10^{36}$\,erg\,s$^{-1}$ at phase $\phi\approx0$. Such a high X-ray luminosity could arise from the photoionized region around the accretion source if the inclination angle of the system is low or if the size of the donor is too small to fully cover the photoionization zone. However, M33\,X-7 is an eclipsing system with a high inclination angle ($\approx75\degr$) and the donor has a large radius over-filling its Roche-lobe. Typically in Roche-lobe overflow systems, due to the high X-ray luminosity, the stellar wind is interrupted by the photoionization. Thus, the wind can only develop in the X-ray shadow behind the OB supergiant, a so-called ``shadow wind'' region \citep[e.g.,][]{Blondin1994,Kaper2005}. In M33\,X-7, the strong X-ray luminosity near phase $\phi\approx0$ suggests the presence of a significant ionization zone in the line of sight of the donor with the shadow wind region being very limited. Such an extended Str\"{o}mgren zone could be formed due to the scattering of X-rays by ions in the stellar wind. As a result, the absorption and emission components of UV wind lines are diminished. By incorporating such a high amount of X-rays we are able to reproduce the observed UV spectra at the eclipse as shown in Fig.\,\ref{fig:m33x7_095}.

As discussed in Sect.\,\ref{sec:xmm}, we estimated the X-ray luminosity of the system at phase $\phi=0.33$ and 0.5 to be around $L_{\mathrm{x}}=10^{38}$\,erg\,s$^{-1}$.
Hence, by including X-rays with $L_{\mathrm{x}}\sim 10^{38}$\,erg\,s$^{-1}$ and adopting the same mass-loss rate ($\log \dot{M}\sim -6.2$), our model SED and spectra are in agreement with the \xmm\ and \hst\ observations at phase $\phi=0.33$ and 0.53 (see Figs.\,\ref{fig:xmodel}, \ref{fig:m33x7_053}, and \,\ref{fig:m33x7_033}). Since the X-ray and UV spectra at phase 0.33 and phase 0.53 are similar and the UV wind lines in these phases suffer large reductions in the absorption strength at all velocities, we could not find any evidence for an accretion wake or focused wind stream toward the BH in this system. 

\begin{figure}
	\centering
	\includegraphics[width=0.47 \textwidth,trim={0 0cm 0 1cm},clip]{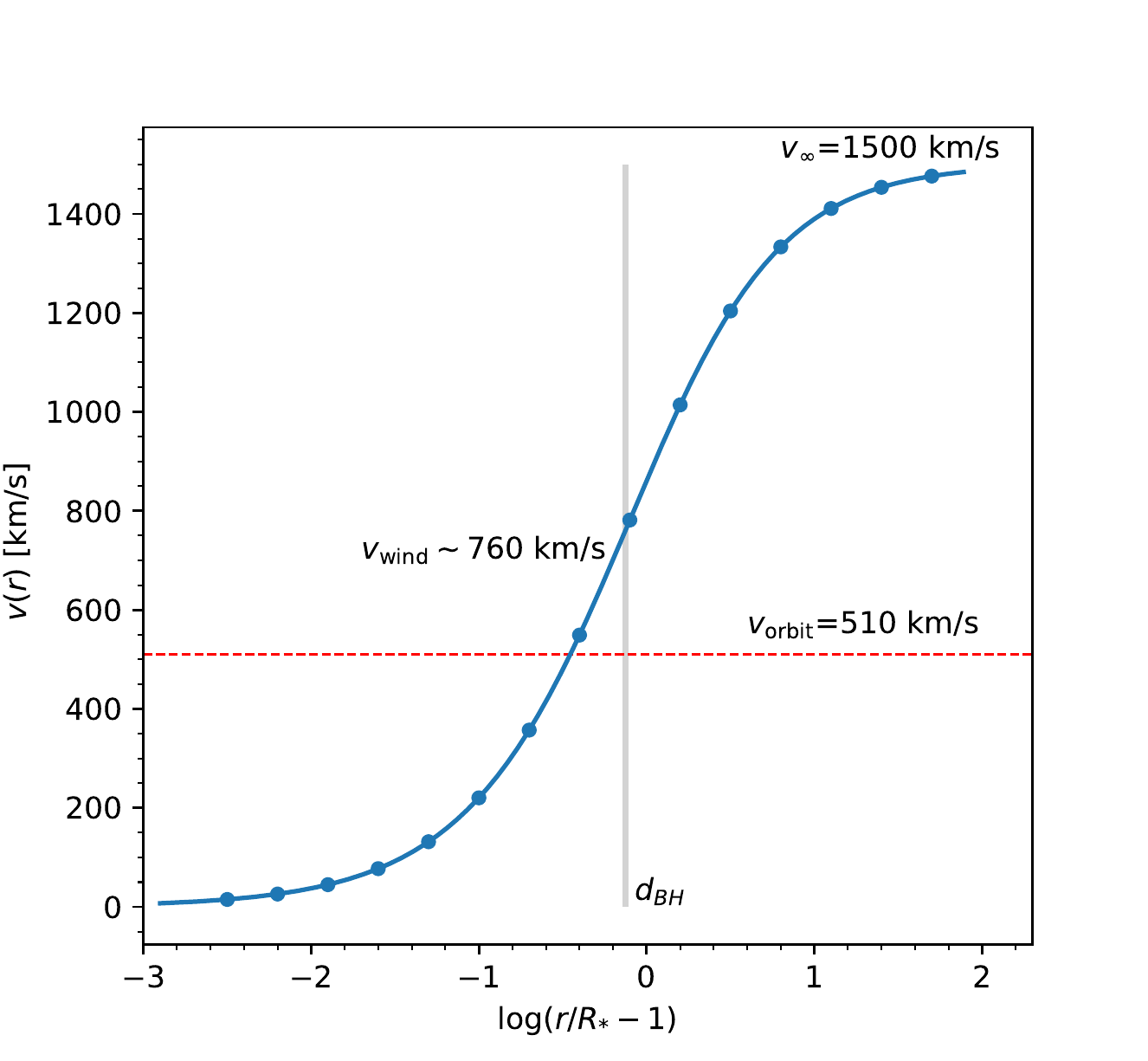}
	\caption{Velocity field for M33\,X-7 donor adopting $\beta=0.8$. The location of BH and the corresponding wind velocity and orbital velocity are marked. }
	\label{velo}
\end{figure}

HMXBs with high X-ray luminosities such as SMC\,X-1 and LMC\,X-4  (OB supergiant + neutron star) show similar orbital-dependent variations in the UV spectra. They show a similarly high X-ray luminosity around deep eclipse ($\sim$ 1\% of peak). As a result, their corresponding UV spectra show strong P-Cygni profiles of \ion{O}{vi} and \ion{N}{v} \citep{Vrtilek2005,Sonneborn2007}. These systems are also reported to have a reduction in the strength of \ion{O}{vi} and \ion{N}{v} due to two orders of magnitude higher X-ray luminosity in other phases. This is exactly what is predicted by the models for M33\,X-7 in Fig.\,\ref{fig:xmodel}. Unfortunately, the UV spectra of M33\,X-7 do not cover these \ion{O}{vi} and \ion{N}{v} lines. In conclusion, our models are capable of explaining the UV wind line variation of high luminosity HMXBs.

Moreover, we varied $\dot{M}$ in models including X-rays to check its impact on the emergent spectral profiles. However, these results either over or under-predict the strength of the UV wind lines compared to the observed spectra.
By fixing mass-loss rate and wind velocity, we have also tested models with different $\beta$ values. The standard $\beta=0.8$ gives the best-fit model whereas higher $\beta$ values could not reproduce the observed UV line profiles. A low $\beta$ value implies a strong acceleration of the wind, hence the wind velocity near the BH (at 1.75$R_\ast$) could reach up to 760\,km\,s$^{-1}$ (see Fig.\,\ref{velo}). In comparison, the orbital velocity ($2\pi a/P$) at this radius is only 510\,km\,s$^{-1}$. However, the actual wind velocity near the BH might very well be comparable to the orbital velocity or even lower as we assume a prespecified wind velocity field in this work. In this approach, the effect of the X-ray illumination from the BH on the spectrum formation is taken into account, but there is no immediate coupling of the velocity field to the accompanying lower radiative acceleration derived in Sect.\,\ref{sec:winddriving}.

In conclusion, we prefer our final best-fit solutions to M33 X-7 using X-ray irradiated models over other possible solutions (see Figs.\,\ref{fig:pcygni_095} and\,\ref{fig:pcygni_053}) since they simultaneously reproduce the X-ray, UV, and optical observations. Moreover, we are able to account for phase-dependent variations in the spectra with the same wind parameters. The wind parameters of our best possible solution are listed in Table.\,\ref{table:parameters}.



\subsection{\heii emission}

\begin{figure}
    \centering
    \includegraphics[width=0.47\textwidth]{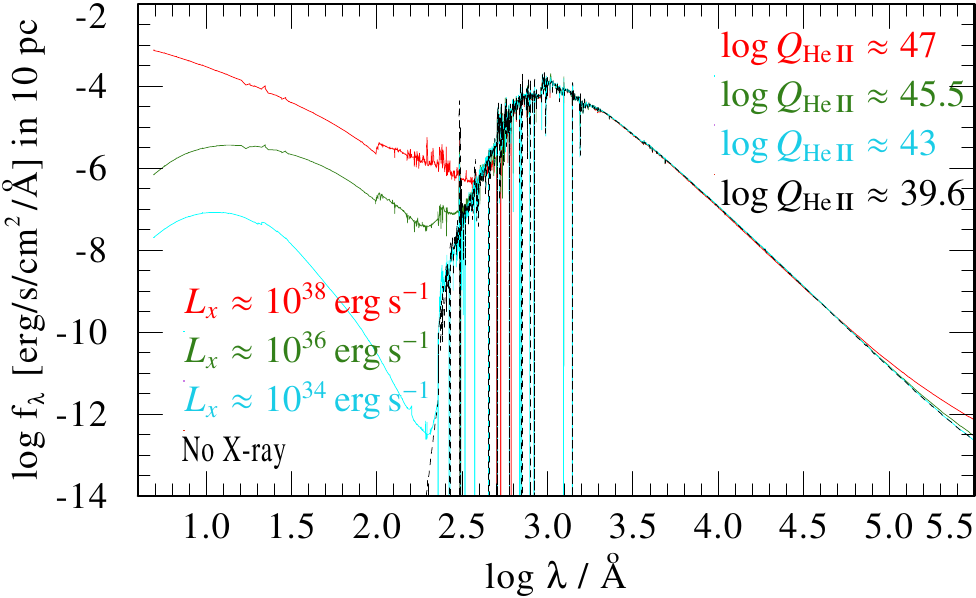}
    \caption{Spectral energy distribution for models with different X-ray luminosities. The corresponding number of \heii ionizing photons are displayed in the legend. }
    \label{fig:xmodelHeii}
\end{figure}

One of the main sources of the \heii emission seen in high redshift galaxies is thought to be HMXBs. Strong X-ray photoionization in HMXBs causes an excess flux beyond the He ionizing edge. This is demonstrated in our model SEDs in Fig.\,\ref{fig:xmodelHeii}. 
The number of \heii ionizing photon flux provided by the O star donor model alone (black lines in Fig.\,\ref{fig:xmodelHeii}) is substantially lower than those  irradiated with X-rays. The \heii ionizing photon flux generated by M33\,X-7 is found to be approximately seven orders of magnitude higher than that from a normal O star of the same spectral type. 

\cite{Maggi2011} have reported nebular\,\heii 4686\,\AA\ emission in the region. They inferred the luminosity of the line to be 
$\sim 2 \times 10^{35}$\,erg\,s$^{-1}$. Using a basic photoionization equation from \cite{Osterbrock2006},  we calculate the\,\heii photon flux to be $Q_{\mathrm{He\,\textsc{ii}}}=  1.02 \times 10^{48} \frac{L(\mathrm{He\,\textsc{ii}\, 4686})}{10^{36}\mathrm{erg\,s^{-1}}} = 2\times 10^{47} \mathrm{ph\,s^{-1}} $. This is in good agreement with our model predictions for M33\,X-7 considering an X-ray irradiation of $L_{\mathrm{x}}=10^{38}$\,erg\,s$^{-1}$ out of eclipse.

\begin{figure}
	\centering
	\includegraphics[width=0.48\textwidth,trim={0 0cm 11.5cm 6.5cm},clip]{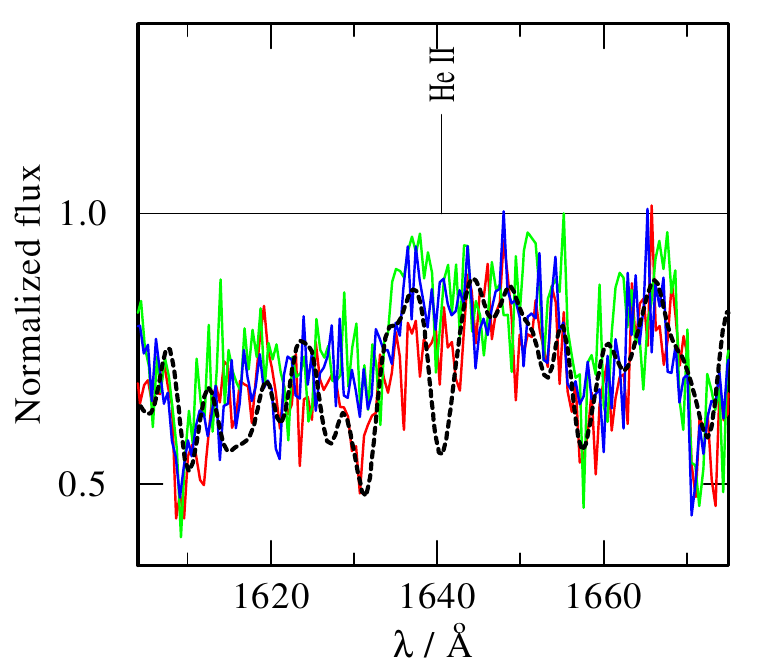} 
	\caption{Comparison of UV spectra of M33-X7 around \heii 1640\,\AA\ taken during different orbital phases. Line colors indicate different orbital phases as given in Fig\,\ref{fig:m33x7_phases}. The dashed black line represent our best fit model including  $L_{\mathrm{x}}=10^{38}$\,erg\,s$^{-1}$, corresponding to $\phi=0.53$. }
	\label{fig:heiiemission}
\end{figure} 

Interestingly, the UV spectra does not show an absorption line at \heii\,1640\,\AA\, despite this being predicted by the model (see Fig.\,\ref{fig:heiiemission}). This absence of absorption suggests that there is possibly an  emission component filling the line. This feature is also variable with orbital phases, with the maximum ``emission'' occuring when  BH is in our line of sight (green and blue lines in Fig.\,\ref{fig:heiiemission}). Such\,\heii emission features are not expected for a typical late O giant star. Although we account for a maximum X-ray luminosity of $\log(L_\text{X}/\text{erg}\,\text{s}^{-1})\approx 38$
 in the model which predicts\,\heii ionizing photons as high as 10$^{47}$\,s$^{-1}$, we were unable to reproduce these observed emission features. Broad\,\heii\,$\lambda$1640 emission features were detected in another HMXB HD\,153919 (alias: 4U1700-37) with a neutron star companion \citep{Kaper1990}. These lines are expected to originate from EUV emission lines close to 256\AA, which is Raman scattered by\,\heii ions in the stellar wind. \cite{Kaper1990} show that these emission lines originate close to the compact object, causing the orbital modulation of the Raman scattered lines. However, HD\,153919 also shows a wind-instrinsic P-Cygni profile in \heii\,$\lambda$1640 \citep{Hainich2020}, which never appear in our M33\,X-7 UV spectra, even at X-ray eclipse.

\section{Accretion rate}
\label{Sect:accretion}

Accretion of matter by the compact object is the most efficient way of producing X-rays. 
In wind-fed supergiants, X-ray accretion is powered by the stellar wind of the donor. 
Accretion of matter onto a star moving through a medium was first described by \cite{Bondi1944} and applied to HMXBs by \cite{Davidson1973}. The mass accretion rate via the Bondi-Hoyle formula can be written as follows:
\begin{equation}\label{mc}
	\dot{M}_\mathrm{acc} =\pi	R_\mathrm{acc}^{2} \rho \varv_{\mathrm{rel}}
\end{equation}
Here $\rho$ is the wind density near the BH and $R_\mathrm{acc}$ is the accretion radius.

In an HMXB, the compact object has a
velocity relative to the stellar wind flow:
\begin{equation}\label{v}
\varv_{\mathrm{rel}}^{2} = \varv_{\mathrm{wind}}^{2} + [\varv_{\mathrm{orbit}} - (R_\ast/d_{\mathrm{BH}})\,\varv_{\mathrm{rot}}]^{2}
\end{equation} Here  $\varv_{\mathrm{wind}}$ and $\varv_{\mathrm{orbit}}$ are the wind velocity and orbital velocity near the BH which are 760\,km\,s$^{-1}$ and 513\,km\,s$^{-1}$ respectively. The eccentricity of the orbit is ignored in this case because it is essentially zero for this system.
Since the O star is rotating fast while the wind is leaving its surface, there is an additional tangential component to its velocity. Plugging in our derived values from the analysis, the estimated velocity is $\varv_{\mathrm{rel}} \approx 845$\,km\,s$^{-1}$.  The derived relative velocity is comparable to $0.56\,\varv_\infty $. 
Stellar wind matter is accreted onto the compact object if it approaches within an accretion radius, which can be approximated to
\begin{equation}\label{R}
		R_\mathrm{acc}= \dfrac{2GM_\mathrm{BH}}{\varv_{\mathrm{rel}}^{2}}
\end{equation}
Using the velocity calculated above, the accretion radius in M33\,X-7 is $\sim 4\times 10^{6}$\,km = $5.7R_\odot$. This is roughly $0.5R_\ast$ from the photosphere of the donor star.
The density of the wind at the distance of the BH can be approximated to 
\begin{align}
	\rho = \dfrac{\dot{M}_\mathrm{donor}}{4\pi d_\mathrm{BH}^{2}  \varv_{\mathrm{wind}} }
\end{align}

By substituting the derived wind parameters, we get a density of $\rho \approx 6.7\times 10^{-15} \mathrm{g\,cm^{-3}}$, and based on Eq.\,\eqref{mc}, we get $\dot{M}_\mathrm{acc}=2.8\times 10^{17} \mathrm{g\,s^{-1}}$.
The expected X-ray luminosity can then be estimated using the relation $L_{X}=\eta L_{\mathrm{acc}} $ with the accretion luminosity 
\begin{equation}\label{L}
L_{\mathrm{acc}} = \dfrac{GM_\mathrm{BH}\dot{M}_\mathrm{acc}}{R_\mathrm{BH}} 
\end{equation}
 and $\eta$ being an accretion efficiency parameter. 
 
In BHs, the accretion disk luminosity will be created while the material falls in from the outermost radius to the innermost stable circular orbit. For a nonrotating BH, the inner edge of the accretion disk should be at three times the Schwarzschild radius $R_\mathrm{sch}= \dfrac{2GM_\mathrm{BH}}{c^{2}}$. For a BH of $11 M_{\odot}$, the accretion radius would be around 32\,km.  For a rapidly spinning BH, the inner edge of the accretion disk can be considerably closer to the event horizon than for a nonrotating BH. The accretion efficiency would be in the range of $\sim6-40\%$. Therefore, the accretion luminosity can be in the range $1-8\times 10^{37} \mathrm{erg\,s^{-1}}$. Previous studies have estimated the spin of M33\,X-7 to be 0.84 \citep{Liu2008,Liu2010}. However, the considerable reduction in the mass of the BH can lower the spin parameter.
Considering the effect of the BH mass on the spin parameter \citep[see Fig.\,3 in][]{Liu2008}, the updated spin of M33\,X-7 is around $a_\ast \approx0.6$. Therefore, the corresponding accretion luminosity is only $\sim 4\times 10^{37} \mathrm{erg\,s^{-1}}$. 

The estimated accretion luminosity corresponds approximately to the observed X-ray luminosity for M33\,X-7 at eclipse  \citep{Pietsch2006,Pietsch2004} but is a factor of four lower than our estimated values for the other two phases out of eclipse (Sect.\,\ref{sec:xmm}). In comparison, the Eddington luminosity of the system is $\sim 1.4\times 10^{39} \mathrm{erg\,s^{-1}}$. So the maximum observed luminosity is only $\approx12\%$ of the Eddington value, which excludes super-Eddington-accretion.

Possible explanations for the observed high X-ray luminosity would be excess mass transfer due to classical Roche-lobe overflow (RLOF) in the system or so-called wind-RLOF, where the radius at which the wind is accelerated is comparable to the Roche-lobe radius so that the wind material fills the donor's Roche-lobe and is transferred to the compact companion through the inner Lagrangian point \citep{Mohamed2007}. The different modes of mass supply likely yield differences in the observed X-ray luminosity of X-ray binaries. Typically, HMXBs which accrete only via stellar winds show a constant X-ray radiation with luminosities in the range $10^{35-37}$\,erg\,s$^{-1}$, while systems associated with RLOF have higher luminosities ($L_{\mathrm{x}}\gtrsim 10^{38}$\,erg\,s$^{-1}$).

In M33\,X-7, the O supergiant companion is in a stage where it is already filling the traditionally defined Roche lobe. However, the presence of a radiative acceleration $\Gamma_\text{rad}$ can significantly alter the Roche potential \citep[e.g.][]{Howarth1997,Dermine+2009}. To get a better idea about this, we can use our insights gained from the radiative acceleration study in Sect.\,\ref{sec:winddriving}. As discussed in \citet{Dermine+2009}, there is no longer a Roche lobe for the donor star in regions with $\Gamma_\text{rad} > 1$. We know from UV spectroscopy that this must be the case for M33\,X-7, at least in some region not facing the BH. Hence, some of the material will directly escape from the system and is not re-directed toward the BH. For $\Gamma_\text{rad} > 1$, there is no longer a clearly defined Roche lobe and thus the  destinction between wind capturing and RLOF actually vanishes.  

For M33 X-7, the situation is likely even more complex. Even the concept of wind-RLOF \citep{Mohamed2007} does not really apply here as it would require $\Gamma_\text{rad} \lesssim 1$ to define a Roche lobe that could be filled by a slow wind as we typically find them in late-type stars. Instead, as our acceleration study shows, our donor star likely has a region with $\Gamma_\text{rad} \ll 1$ facing the BH. In this case, there is a definable modified Roche potential with generalized radius, which will be smaller than the one for the unperturbed Roche lobe \citep[see][]{Dermine+2009}. Since the derived donor star parameters for M33 X-7 are already sufficient to yield a star that would fill the standard definition of a Roche Lobe, the further reduction of this criterion for a region towards the BH paves the way for an additional ``easy'' overflow-like mass transfer, that is further boostered by the diminished radiative acceleration of the material due to the X-rays from the BH.  
In contrast to the donor, a meaningful Roche lobe can still be defined for the accretor, which would play a major role in capturing a considerable ($>10\%$) fraction of the wind as indicated in recent hydro-dynamical simulations by \citet{ElMellah2019}. 

Based on our atmosphere modeling, the maximum wind velocity outside eclipse should be $\varv_\infty \approx 1000\,\mathrm{km}\,\mathrm{s}^{-1}$, which is similar to other high-$L_\mathrm{X}$ HMXBs. Hence, we can assume that the wind speed at the distance of the BH is lower than in our eclipse model (cf.\ Fig.\,\ref{velo}), i.e., close to the orbital speed. Subsequently, the accretion luminosity in the Bondi-Hoyle scenario already increases to $\sim 1.4\times 10^{38} \mathrm{erg\,s^{-1}}$, in better agreement with the X-ray observations. With the additional mass overflow in the highly ionized region, we have a plausible ``wind overflow'' scenario for the observed $L_\text{X}$, even taking into account the efficiency factor between $L_{\mathrm{acc}}$ and $L_\text{X}$. While our scenario likely explains the higher accretion luminosity compared to the classical Bondi-Hoyle wind-fed scenario, the condition of $\Gamma_\text{rad} > 1$ in other directions will avoid a standard RLOF where all of the donor's mass loss is redirected toward the BH. Consequently, our derived $L_\text{X}$ is on the lower end compared to systems that are considered to be powered by standard RLOF.

Recent studies of other BH HMXB systems, namely Cyg\,X-1 \citep{Orosz2011,Miller-Jones+2021} and LMC\,X-1 \citep{Orosz2009,Hyde2017}, also suggest that the donor stars in these systems are close to filling their Roche lobe. Yet, a detailed consideration of the radiative acceleration has so far not been done. Depending the the impact of the X-ray illumination of the donor atmosphere, these systems will either have a similar ``wind overflow'' like M33\,X-7 or a standard RLOF in case of strong wind inhibition.

\begin{table*}
	\caption{Comparison of system parameters from single and binary evolution models that best match the derived parameters from spectroscopic analysis of M33 X-7}
	\label{table:evolutionparameters}
	\centering
	\renewcommand{\arraystretch}{1.4}
	\begin{tabular}{l|c|c|c|c}
		\hline 
		\hline
		Parameter                                   &   Single star model    &  Binary  model\,1 & Binary model\,2   &        Empirically derived      \\
		\hline                                          
		$T_{\ast}$ (kK)                             & $31.2^{+2}_{-1.5}$              &  $31.4 $    &    $30.9$     & $31^{+2}_{-2}$         \\
		$\log g_\ast$ (cm/s$^{-2}$)                 & $3.43^{+0.08}_{-0.17}$         &  $3.44 $   &3.34     & $3.4^{+0.2}_{-0.2}$    \\
		$\log L$ ($L_\odot$)                        & $5.52^{+0.07}_{-0.06}$        &  $5.54 $    & 5.54  & $5.54^{+0.1}_{-0.1}$   \\
		$R_\ast$ ($R_\odot$)                        & $19.84^{+3}_{-3.5}$            &  $20$    &20.6       & $20.5^{+2}_{-2}$       \\
		$\log \dot{M}\mathrm{_{wind}}$ ($M_\odot \mathrm{yr}^{-1}$) & $-6.22^{+0.14}_{-0.12}$       &  $-6.35$  &  -6.1  & $-6.2^{+0.1}_{-0.2}$  \\
		$\varv_{\mathrm{rot}}$ (km\,s$^{-1}$)                    & $250^{+13}_{-15}$           &  $30$   & 278      & $280^{+30}_{-30}$      \\
		$X_{\rm He}$ (mass fr.)                      & $0.26^{+0.1}_{-0.1}$        &  $0.28$    &0.3   & $0.395^{+0.1}_{-0.1}$   \\
	    $\log(C/H)+12$             & 7.58              &  7.64    &     7.56    & 7.05          \\
		 $\log(N/H)+12$               & 7.84            &  7.63    &  7.67     & 7.85       \\
		 $\log(O/H)+12$           &8.27               &  8.3      &    8.32   & 8.45        \\
		$P$\,(days) & &4.1 &6.3 & 3.45\\
		$M_\mathrm{O}$ ($M_\odot$)                  & $33.8^{+2.5}_{-2}$            &  $39.4$    &34.2       & $38^{+22}_{-10}$       \\
		$M_\mathrm{BH}$ ($M_\odot$)                 &         &  $\lesssim10.5$   & $\lesssim12.2$    & $11.4^{+3.3}_{-1.7}$   \\
		$d_{\mathrm{BH}}$ ($R_\ast$)                              &          &  $2$   & 2.5   & $1.75^{+0.2}_{-0.1}$   \\
		$R_\mathrm{rl}\,(R_\ast)$                   &       &  $1$    & 1.1  & $0.84^{+0.2}_{-0.1}$   \\
		Age\,(Myr)&$4^{+0.27}_{-0.26}$   &5.76 &5.08&\\
		\hline
	\end{tabular}
\end{table*}

\section{Formation and evolution of the HMXB}

M33 X-7 has been identified as an evolutionary challenge, considering the previously reported very high mass donor star along with a heavy BH in a tight orbit. The updated stellar and binary parameters allow us to gain new insights into the evolution of this system. In the first step, we explore the formation channel of the system starting from a binary with two stars on the zero-age main-sequence (ZAMS) to a BH - O star binary and then follow its path to later evolution.

As a first approach, we compare the updated parameters of the O star donor to that of single-star evolutionary tracks. For this purpose we used tracks provided by \cite{Brott2011} at LMC metallicity along with the Bayesian statistic tool ``The BONN Stellar Astrophysics Interface'' \cite[BONNSAI\footnote{https://www.astro.uni-bonn.de/stars/bonnsai/},][]{schneider2014} . The results from the analysis are summarized in Table\,\ref{table:evolutionparameters}. Single-star evolutionary tracks suggest lower masses for the O star donor given its observed luminosity, i.e., the donor is underluminous compared to a normal single star of the same mass. 
The observed wind mass-loss rate of the donor is in good agreement with the prediction of \cite{Vink2000,Vink2001} which is used in the stellar evolutionary models. The single star models could account for the fast rotation of the star by assuming an initial velocity of 280\,km\,s$^{-1}$. Abundances are also reproduced within the uncertainty limit except for helium.

With our knowledge of the currently observed properties of the M33\,X-7 system, we tried to reconstruct its evolution from the moment the system was comprised of two ZAMS stars through the intermediate phases such as mass transfer from the primary to the secondary, and the formation of the BH, followed by the currently observed HMXB phase and the final evolution of the system.

\begin{figure}
	\centering
	\includegraphics[width=0.5\textwidth,trim={0 0cm 0cm 0}]{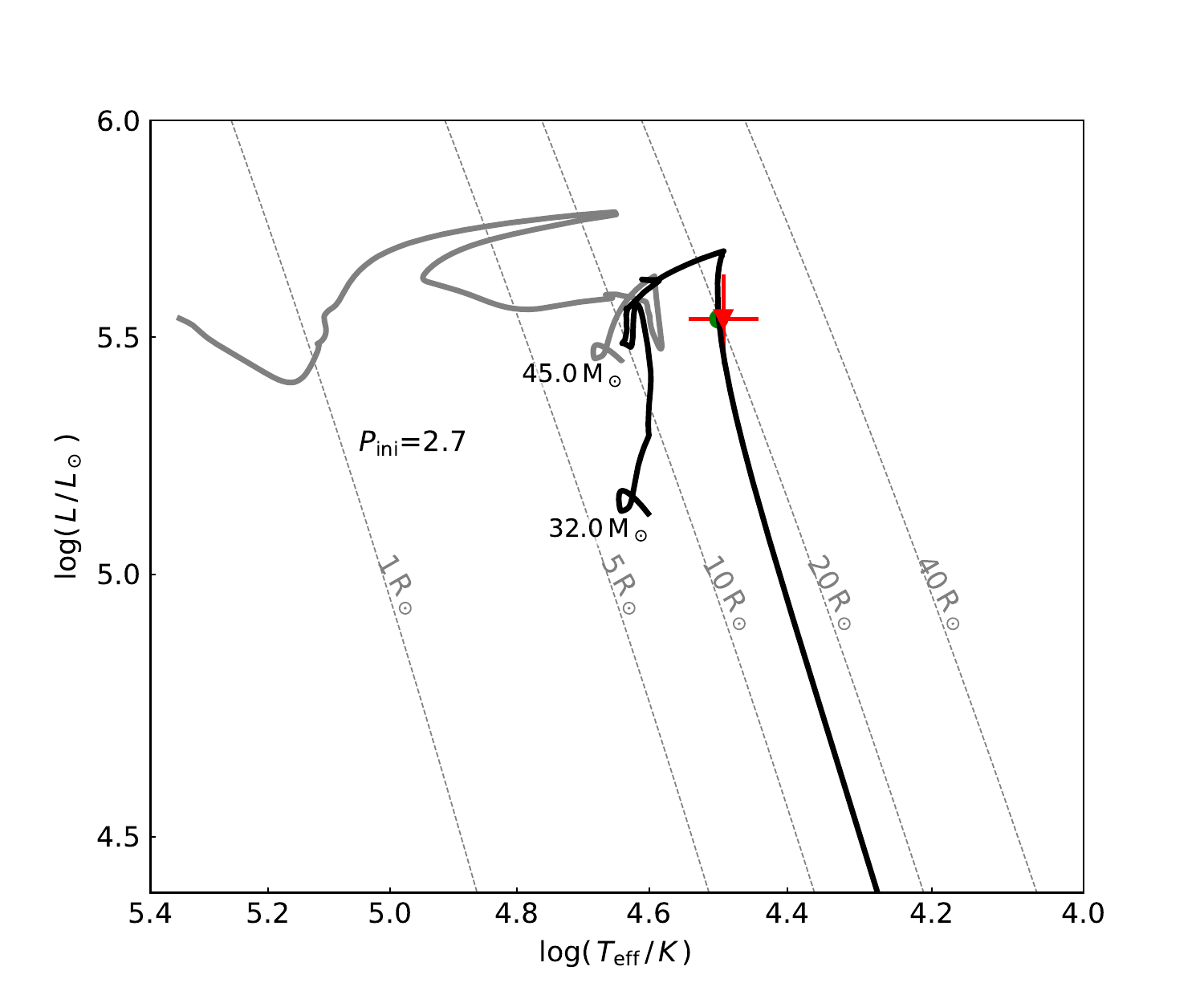}

	\caption{Best matching MESA evolutionary tracks reproducing the HRD position of O-star donor in M33\,X-7. The current position and errors of the O star are taken from the spectroscopic analysis. Initial masses and periods are labeled on the plot.  }
	\label{fig:mesaHRD1}
\end{figure}

\subsection{Modeling the binary system with MESA}
We performed detailed binary evolution calculations to explore the possible evolutionary history of M33\,X-7. The tracks were computed using the stellar evolution code MESA (``Modules for Experiments in Stellar Astrophysics'', Version No.\ 10398) with a physics implementation as described in \cite{Paxton2011,Paxton2013,Paxton2015,Paxton2018}. We adopted LMC metallicity (Z = 0.006) for the calculation, which is in line with the derived metallicities based on the spectroscopic analysis. To make our models comparable to the Brott (2011) LMC models used for the single star evolution in BONNSAI, we modified MESA as described in \cite{Marchant2016}. 

Stellar winds were implemented as in \cite{Brott2011}, with mass-loss rates for hot hydrogen-rich stars modeled as in
\cite{Vink2001}. For stars with surface hydrogen of $X_{\mathrm{H}} < 0.4$, we use the mass-loss rate of \cite{Hamann1995} divided by a factor of 10 to account for clumping \citep{Yoon2010}. We interpolate these two mass-loss rates for surface hydrogen fractions between 0.7 and 0.4. For temperatures below the bi-stability jump, the mass-loss rate was taken as the maximum between the Vink mass-loss rate and that of \cite{Nieuwenhuijzen1990}.  

Convection was modeled using the mixing-length theory \citep{BohmVitense1958} with a mixing-length of $\alpha_{ML}$ = 1.5. Semiconvective mixing is modeled with an efficiency of ${\alpha_\mathrm{SC} = 1}$.

Rotational mixing and angular momentum transport were modeled as diffusive processes following  \cite{Heger2000}, including the effects of Eddington-Sweet circulations, the Goldreich-Schubert-Fricke instability, secular and dynamical shear, and sensitivity to composition gradients. We also take into account the transport of angular momentum due to the Spruit-Tayler dynamo \citep{Spruit2002}. 

Stars are assumed to be synchronized with the orbital period at the ZAMS. The evolution of orbital angular momentum considers the effects of mass loss, gravitational wave radiation, and spin-orbit coupling, as described in \cite{Paxton2015}.

\begin{figure}
	\centering
	\includegraphics[scale=0.4]{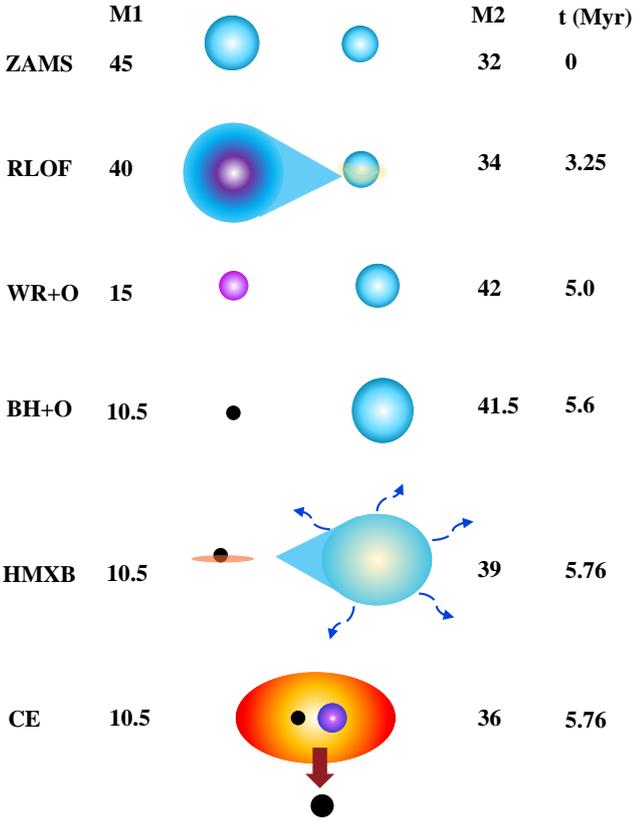} 
	\caption{Illustration of possible evolutionary scenario for M33\,X-7 starting from the ZAMS massive binary to HMXB stage followed by common envelope merging.}
	\label{illu}
\end{figure}

\begin{figure}
	\centering
	\includegraphics[width=0.5\textwidth,trim={0 0cm 0cm 0.9cm},clip]{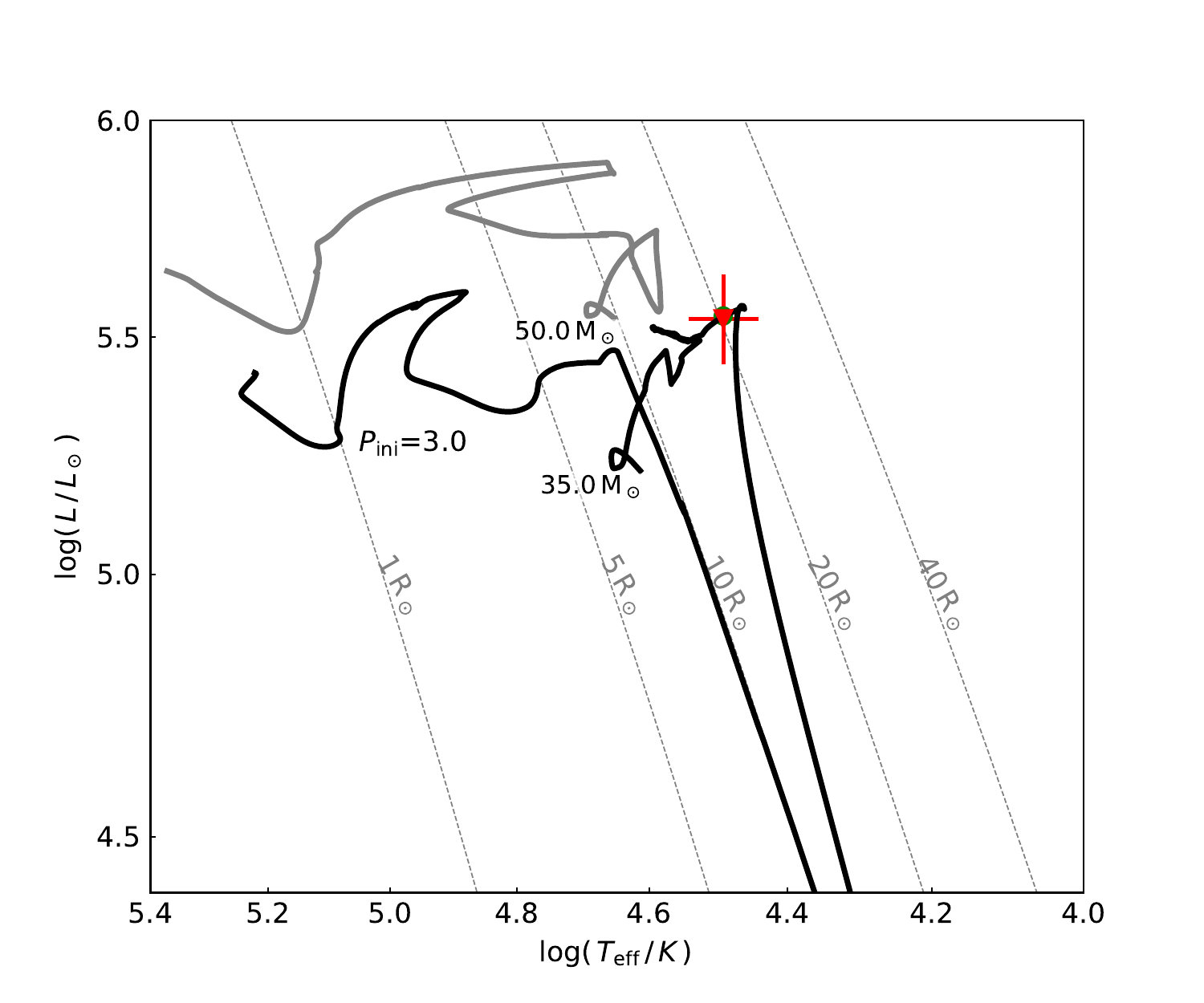}
	\caption{Similar to Fig.\,\ref{fig:mesaHRD2}, but an alternative evolutionary track assuming no tidal synchronization. }
	\label{fig:mesaHRD2}
\end{figure}

\subsection{Evolutionary status of M33\,X-7}
To investigates possible scenarios of M33\,X-7's evolutionary history, we considered progenitor masses in the range of $M_{1}=38-60\,M_\odot$ for the primary, $M_{2}=18-40\,M_\odot$ for the secondary, and initial orbital periods from 1.8 to 5 days. Guided by the mass derived from the spectroscopic analysis of the donor, which is higher than what is expected from evolutionary models of single stars, we searched for binary models where the secondary accretes mass from the BH progenitor. Given the short orbital period, we considered only binary systems that start their life in a tight orbit, hence undergoing a phase of mass transfer during the main sequence phase of the primary. In our model calculation, we allow both stars to evolve up to core carbon depletion. After core carbon depletion, a star is assumed to directly collapse into a BH without losing mass or the orbit receiving a kick due to a possible supernovae explosion.
In this approach, we attribute the final CO mass to the compact companion and treat it as a point mass. Subsequently, we model the evolution of the secondary together with the compact object in the binary system. 

While exploring the parameter space, we discovered that 1) models with high initial masses and short orbital periods are predicted to undergo overflow through the second Lagrangian point L2 during core-H burning and are expected to merge; 2) models with lower mass ratios ($q\lesssim0.5$) and short periods ($P<5$ days) had numerical issues during the fast Case A mass transfer phase; and 3) models with mass ratios going toward unity ($q\gtrsim0.85$) have a reverse mass transfer from the secondary to the primary while both stars are still core-H burning, leading to an unstable situation.
These models terminate before forming a BH + O star binary, and are therefore ruled out, limiting the possible parameter space.
For some models that successfully reach the HMXB stage we found that they form a convective envelope during mass transfer on the compact object, leading to unstable situations followed most likely by a common envelope phase and a merger. We think that these models are unlikely to represent the true nature of this system which is believed to be currently in a stable mass transfer phase. 


Based on our tests, the resulting models exhibit the following properties at the HMXB phase:  $M_{\mathrm{OB}}\approx 30-40\,M_\odot$, $M_{\mathrm{BH}}\approx 8-13\,M_\odot$, and $P\approx 4-12$\,days. Our best possible evolutionary model which matches the current parameters of M33\,X-7 is shown in Fig.\,\ref{fig:mesaHRD1} (Binary model\,1). The parameters derived from this model are compared to the observations in Table\,\ref{table:evolutionparameters}. This model suggests the initial masses of the stars to be $M_{1}=45\,, M_{2}=32\,M_\odot$ with a period of 2.7\,days. Our best model satisfies the current observational properties ($T_\ast$, $\log g$, $\log L$, $R_\ast$) of the donor star. Moreover, the evolutionary masses suggested by the model for the O-star donor and the BH are in good agreement with the observations. 

The initial masses of the M33 X-7 progenitors result from our tests are comparable to typical main-sequence O-stars and are much lower than the predictions from previous studies \citep{Abubekerov2009,Valsecchi2010Natur,deMink2010ASPC, Bogomazov2014ARep}. According to the evolutionary model, the donor is overflowing its Roche lobe and transferring mass to the BH at the currently observed stage as a HMXB. As a result, the total mass-loss rate of the system is an order of magnitude higher in this model, despite the wind mass-loss rate being in agreement with the empirical estimation. This would qualitatively be in line with out findings in Sect.\,\ref{Sect:accretion}, where we conclude an enhanced mass transfer due to the strong X-ray illumination and the perturbance of the Roche potential due to the presence of a donor wind. Of course such a detailed consideration of the wind is beyond the MESA treatment where standard Roche-lobe overflow is assumed instead.

Our suggested binary model\,1 does not reproduce the fast rotation of the secondary star (the current O-star donor). The current period of the system suggested by the model is slightly higher than the observed period. This is mainly because the initial orbit got widened as a result of the strong mass-loss rate of the primary during the assumed Wolf-Rayet phase. When reducing the initial period, we find no stable solution during the first mass-transfer phase. However, adopting a reduced mass-loss rate in the WR stage compared to the standard prescriptions as recently derived for lower metallicities by theoretical calculations \citep[][]{SanderVink2020,Higgins2021} might be able to resolve this conflict. 

Considering the predicted abundances, we find that He is by far not as enriched in our model as derived from our observations. There are major effects that can lead to altered abundances, mixing and losing mass, so that the inner enriched layers are exposed to the observer. As the fast rotation rate of the OB star is not reproduced either, these two findings are likely linked and the evolutionary abundances suffer from missing the additional rotational mixing.
 
Figure\,\ref{illu} illustrates possible evolutionary sequences of M33\,X-7 based on the results from our best-match model. The progenitor comprises of a primary with $45\,M_\odot$ (BH progenitor) and a secondary with $32\,M_\odot$ (O-star donor progenitor). At this point, both stars are on the main sequence. During the first $\approx 3$\,Myr the system remains detached. After core-hydrogen exhaustion, the primary leaves the main sequence and starts to expand quickly. When its radius approaches the Roche lobe, a rapid phase of mass transfer initiates to the less massive secondary which still resides on the main sequence. The secondary accretes most of the transferred mass and acquires large angular momentum so that it can spin up to critical velocity. During this phase, the orbit shrinks to 2.5 days. Even though the secondary spun up during the mass transfer, the rotation rate drops quickly to the earlier values after this stage due to tidal locking. After the rapid interaction the system experiences episodes of slow mass transfer until $\approx4$\,Myr. The mass transfer ends when the primary’s hydrogen envelope is partially stripped, with only $\approx30\,M_\odot$ left. The primary will evolve to the core-He burning phase, transitioning to a Wolf-Rayet star with a strong stellar wind. At this point, the mass of the He core would be around  $15\,M_\odot$. Due to its strong mass loss, the orbital period would increase to 5.5 days. The Wolf-Rayet phase will last for $0.6$\,Myr. The remaining $He$ core of the primary eventually collapses into a BH at 5.6 Myr.

At around $5.76$\,Myr, the scenario described above forms a HMXB, resembling the properties of M33\,X-7. The total duration of this phase lasts for $\lesssim0.2$\,Myr, implying that this stage has a reasonable probability to be observed. At this stage, similar to our observations, the O-star donor has already filled up its Roche lobe, and the BH is accreting this material (though see Sect.\,\ref{Sect:accretion}).

According to our model, the mass-transfer will become unstable in a few 10,000 years and the system will merge. However, a more fine-tuned model (e.g., with different mixing) might lead to a stable solution. Nonetheless, since the mass ratio is large and the orbital period is short, it does not seem to be unreasonable that the system will eventually merge\footnote{\revise{However, the formation of a supercritical accretion disk around the BH may stabilize the mass transfer in a way similar to the case of SS433 with similar mass ratio, see for e.g., \citet{Cherepashchuk2021}}}. Under this assumption, our system will not be a progenitor of a double-BH binary. Instead, the BH and the He core will form a common envelope with an in-spiral of the BH eventually resulting in a merger, leading to the formation of a more massive BH. This could be accompanied by a $\gamma$-ray burst.

As a slightly different alternative we show a second binary model (Binary model\,2) that also reproduces the observed parameters quite well while having a different fate. The results are shown in Table\,\ref{table:evolutionparameters} and the corresponding evolutionary tracks are plotted in Fig.\,\ref{fig:mesaHRD2}. In this scenario, we have discarded our assumption of tidal synchronization in the binary system. However, it is most likely that the short period binaries would evolve into tidal locking very soon, independent of the initial stellar spins. In binary model\,2, we assume an initial rotation velocity of 300\,km\,s$^{-1}$ for both stars.  
Due to the change in our assumptions on rotation, now the secondary is only accreting a small fraction of mass during the first mass transfer. Contrary to the earlier model, the rotation rate of the secondary reduces only gradually after the spin up. Hence, this model resembles the observed fast rotation of the O-star donor at the current HMXB phase. Subsequently, the primary quickly proceeds to the Wolf-Rayet phase at around 4.5\,Myr with a He-core mass of $\approx20\,M_\odot$. At 5\,Myr the primary collapses to a $12\,M_\odot$ BH. At this stage, the orbit has already widened due to the strong wind mass loss. The HMXB phase in this model happens earlier than in the previous model. However, the period of the system is almost a factor two higher than that of M33\,X-7. In addition, the mass of the donor is lower whereas the BH mass is slightly higher compared to our empirical estimations. This translates to a lower mass ratio for the system ($q\approx2.8$). Because of this slight decrease in the mass ratio and increase in the orbital period, the second model predicts a stable mass transfer phase after the donor fills its Roche lobe at 5.4\,Myr, showing the complexity of stellar evolutionary modeling. At this point, the orbital period shrinks down to two days. Subsequently, the secondary starts to contract and evolves further as a Wolf-Rayet star.  Throughout these final phases, the primary BH accretes and gains around $1\,M_\odot$ via stellar winds.
 At 6.3\,Myr, the secondary will reach core carbon burning exhaustion. The final system consists of a BH-BH binary with 13 and $9\,M_\odot$ in a four-day orbit. Based on their large separation, these BHs will not merge in a Hubble time.

\paragraph{On the elusive detection of M33\,X-7-like systems:} Although the initial masses of the M33 X-7 progenitors are comparable to that of typical O-star binaries, such BH HMXB systems are rare. Using binary evolutionary model calculations, we investigated the formation of OB + BH binaries with a range of masses and periods. According to these models, the duration of the OB + BH system is only $\approx 2 - 5$\% of the total lifetime of the binary before both stars become compact objects. Moreover, such a system would be observed as X-ray binaries only when an accretion disk forms around the BH. The accretion disk formation criterion for wind-fed systems mainly depends on the wind and orbital velocity near the BH radius and on the mass ratio \citep{Illarionov1975,Shapiro1976}. Thus, BH HMXBs are mostly limited to short-period systems with donors evolved from the main sequence. With the same observed parameters as of M33 X-7, a system would not be detectable in X-rays if the orbital period is $>15$ days. Besides, if the donor is a main-sequence star with a smaller radius and a higher terminal wind velocity than in M33 X-7, the accretion disk will only form when the orbital period is less than its current value.

\section{Conclusions}
\label{sect:conclusions}
We present a detailed spectroscopic analysis of the only known eclipsing BH HMXB system, M33\,X-7.
For this study we obtained the first UV spectra of the system accompanied by X-ray observations. The observations were carried out at three key orbital phases, tracing the BH eclipse, egress, and inferior conjunction using the \hst. At the latter two phases, X-ray observations were obtained simultaneously using \xmm. These phase-resolved observations shed light on the interaction of the stellar wind with the BH. The UV resonance lines show the Hatchett-McCray effect with a large reduction in absorption strength when the BH is in the foreground due to the strong X-ray ionization.
We performed a detailed analysis of X-ray, UV, and archival optical spectra using stellar atmosphere models, arriving at the following conclusions:
\begin{itemize}
    \item The donor is much cooler and dimmer than previously predicted. The mass-accreted O star shows signs of He enrichment and fast rotation.
    
    \item \revise{Our detailed analysis considering multi-wavelength data suggests a large reduction in the mass of the system compared to previous results. The mass of the donor star is reduced from $70\,M_\odot$ to $38\,M_\odot$.  Correspondingly, the BH mass changes from $15.6\,M_\odot$ to $11.4\,M_\odot$ and spin of the BH decreases to $a_\ast \approx0.6$}.
    
    \item Based on the newly derived parameters of the donor and mass of the BH, the orbital separation has come down to $35\,R_\odot$. The comparison of the radius of the star with a standard Roche-lobe approximation suggests an overfilling factor of $f\approx1.2$. 
    
    \item The strong X-ray radiation of M33 X-7 ionizes much of the wind and leaves a large Str\"{o}mgren zone. These extreme ionization conditions prevent the formation of a normal radiatively driven wind from significant parts of the donor. Our one-dimensional calculations confirm that the photoionization by the X-ray radiation can significantly change the ionization structure and diminish the wind accelerations. Our investigations on wind driving and the impact X-ray irradiation in M33 X-7 can also be applied to other high-luminosity HMXB systems in general. 

    \item We estimated the wind parameters of the donor based on the UV spectra taken at X-ray eclipse. The derived mass-loss rate is in good agreement with the theoretical prediction assuming a depth-dependent microclumping. By incorporating  X-rays corresponding to the observations at different orbital phases, we were able to reproduce the observations at three different phases including eclipse with the same mass-loss rate.

    \item The observed X-ray luminosity derived from our analysis is $\sim 2 \times 10^{38} \mathrm{erg\,s^{-1}} $, which is about 12\% of the BH Eddington luminosity. 
    The accretion luminosity derived using the Bondi-Hoyle calculation is found to be much lower than the observed X-ray luminosity. Fostered by the wind inhibition due to the X-ray photoionization, we instead have a ``wind overflow'' scenario where the BH not only captures a part of the wind, but it is further fed by an additional mass overflow along the inner Lagrangian point. In other directions, the presence of a significant stellar wind significantly perturbs the Roche potential, allowing material to escape from the system. As a consequence, the classical distinction between wind-fed and Roche-lobe overflow systems becomes meaningless for our system. The observed
    X-ray luminosity is exactly in line with such a ``wind overflow'' scenario. 

    
    \item  The X-rays contribute to the formation of the \heii\ emission region around the system emitting $\sim 10^{47} \mathrm{ph\,s^{-1} } $.

\end{itemize}

Taking the parameters of  M33\,X-7 from the observations into account, we attempted to trace its evolution using binary evolution models. We modeled the system from a ZAMS binary system until its current state. In addition, we proposed two possible scenarios for its future evolution. 
The system is transitioning toward an unstable mass transfer phase, resulting in a common envelope with a deep spiraling of the BH to the envelope of the massive donor. Since the mass ratio is q$\gtrsim$3.3 and the period is short, the common envelope phase is likely to result in the merger of two companions resulting in a heavier BH.
\begin{acknowledgements}
The authors are grateful to the referee for the detailed and constrictive 
report which helped to improve the paper, and also for providing the optical spectroscopic data.
We thank O. Kargaltsev for the useful discussions. We thank the {\em Swift} telescope team for approving our ToO request.  
This work made use of data supplied by the UK {\em Swift} Science Data Centre 
at the University of Leicester.
VR acknowledges support by the Deutsches Zentrum für Luft- und Raumfahrt (DLR) under grant 50 OR 1912. VR and AACS acknowledge support by the Deutsche Forschungsgemeinschaft (DFG - German Research Foundation) in the form of an Emmy Noether Research Group (grant number SA4064/1-1, PI Sander).
DP acknowledges financial support by the Deutsches Zentrum für Luft und Raumfahrt (DLR) grant FKZ 50 OR 2005.
This publication has benefited from discussions in a team meeting sponsored by the International Space Science Institute at
Bern, Switzerland.
This research made use of the VizieR catalog access tool, CDS,  Strasbourg, France. The original description of the VizieR service was published in A\&AS 143, 23. 
\end{acknowledgements}

\bibliographystyle{aa}
\bibliography{ref}

\appendix

\section{Appendix}

\begin{figure*}[!htb]

\caption{Comparison of synthetic spectra in the blue optical range obtained for models with different stellar temperatures and surface gravities. The mean optical spectrum of M33 X-7 \citep{Orosz2007} is shown as a black solid line. The deep cores in some of the Balmer lines are likely due to an over-subtraction of nebular
emission  \citep[see supplementary information in][]{Orosz2007}. In contrast to the other \heii lines, \heii 4686\AA\, appears as a weak emission in the observation. This line might be filled up with nonstellar emission, e.g.\ from the accretion disk. }
\label{fig:opt}
\vspace{0.5cm}
\includegraphics[width=\textwidth]{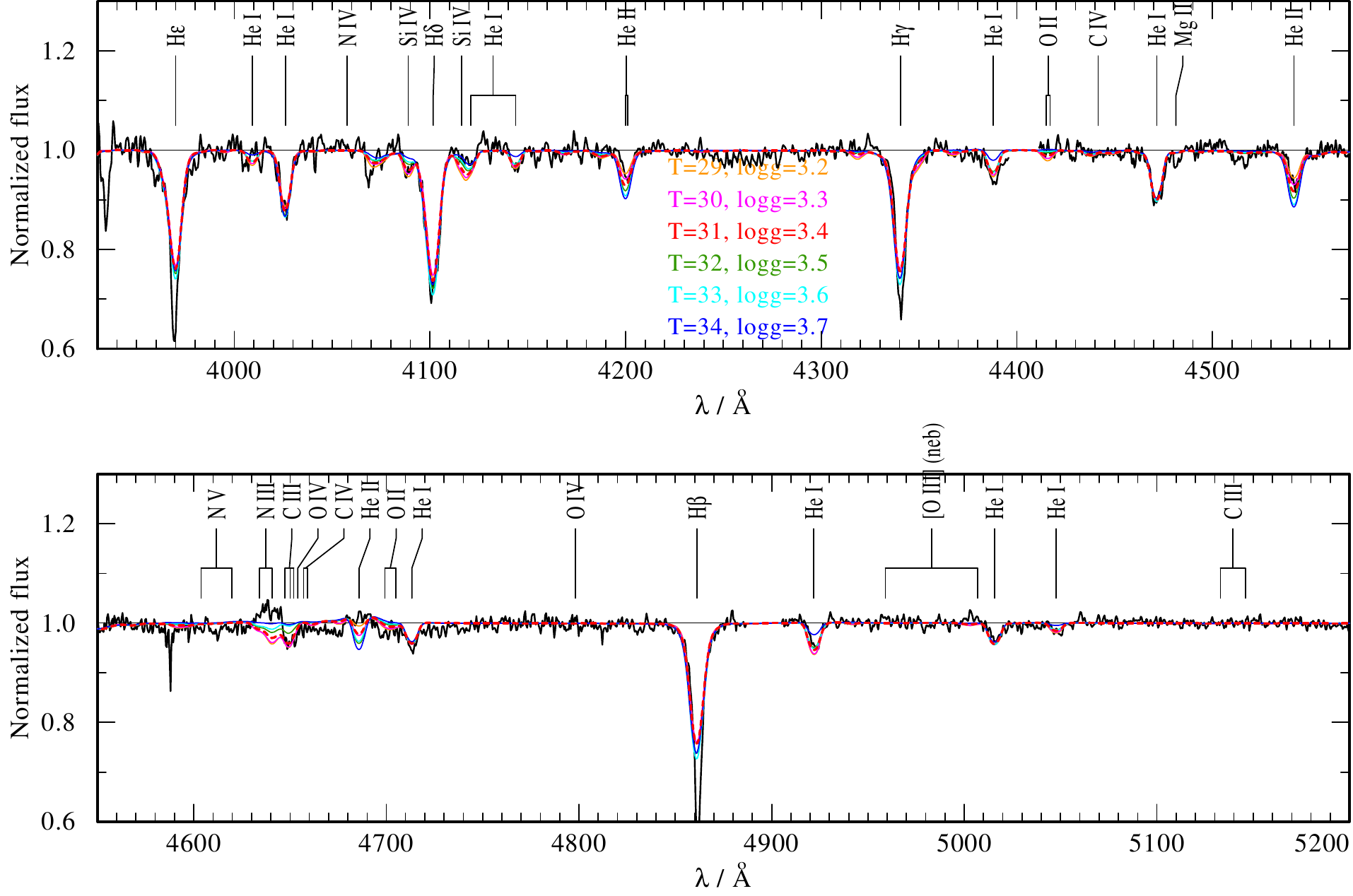}
\end{figure*}
\begin{figure*}[!htb]
\caption{Same as figure \ref{fig:opt}, but zooming on to H$\delta$ comparing models with different surface gravities.}
\label{fig:opt3}
\vspace{0.5cm}
\includegraphics[width=0.45\textwidth]{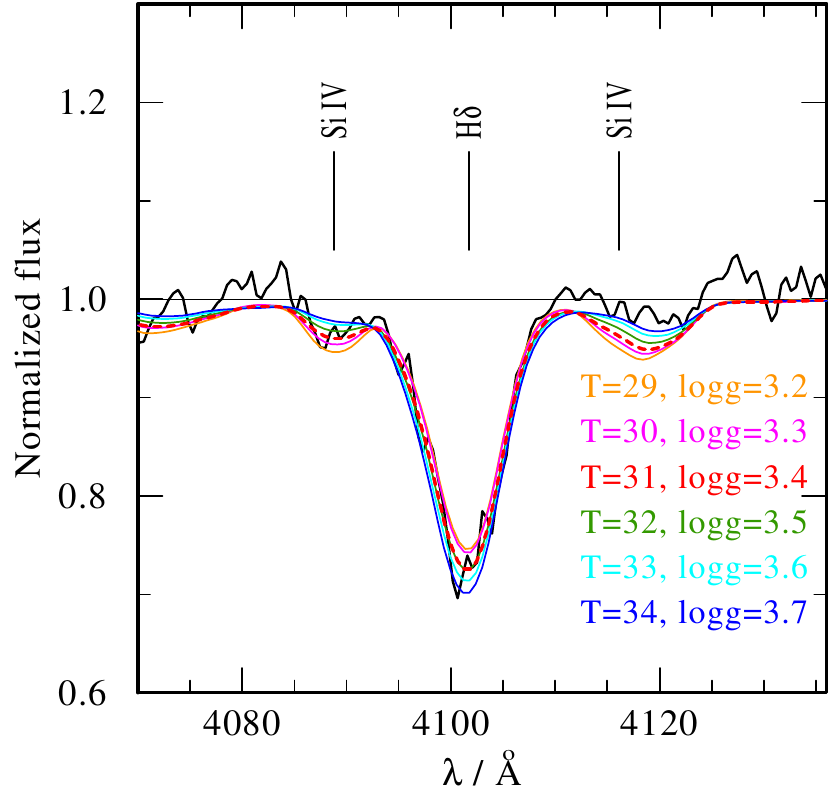} 
\includegraphics[width=0.45\textwidth]{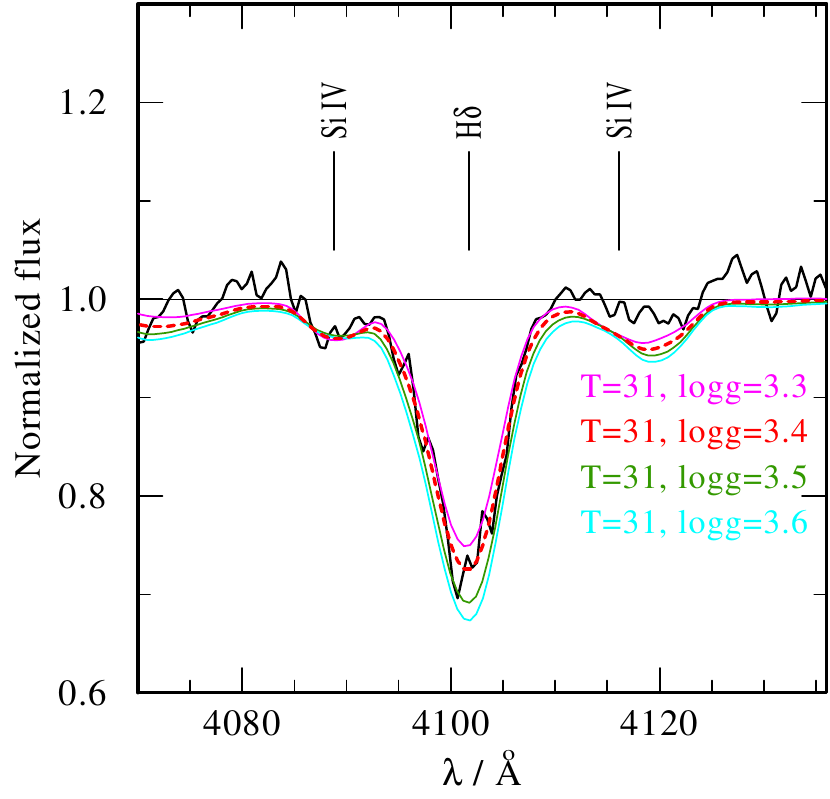}
\end{figure*} 

\begin{figure*}[!htb]

\caption{Same as figure \ref{fig:opt}, but zooming on to multiple \hei\ and \heii\ lines}
\label{fig:opt2}
\vspace{0.5cm}
\includegraphics[width=0.9\textwidth,trim={0cm 0cm 1cm 0cm}]{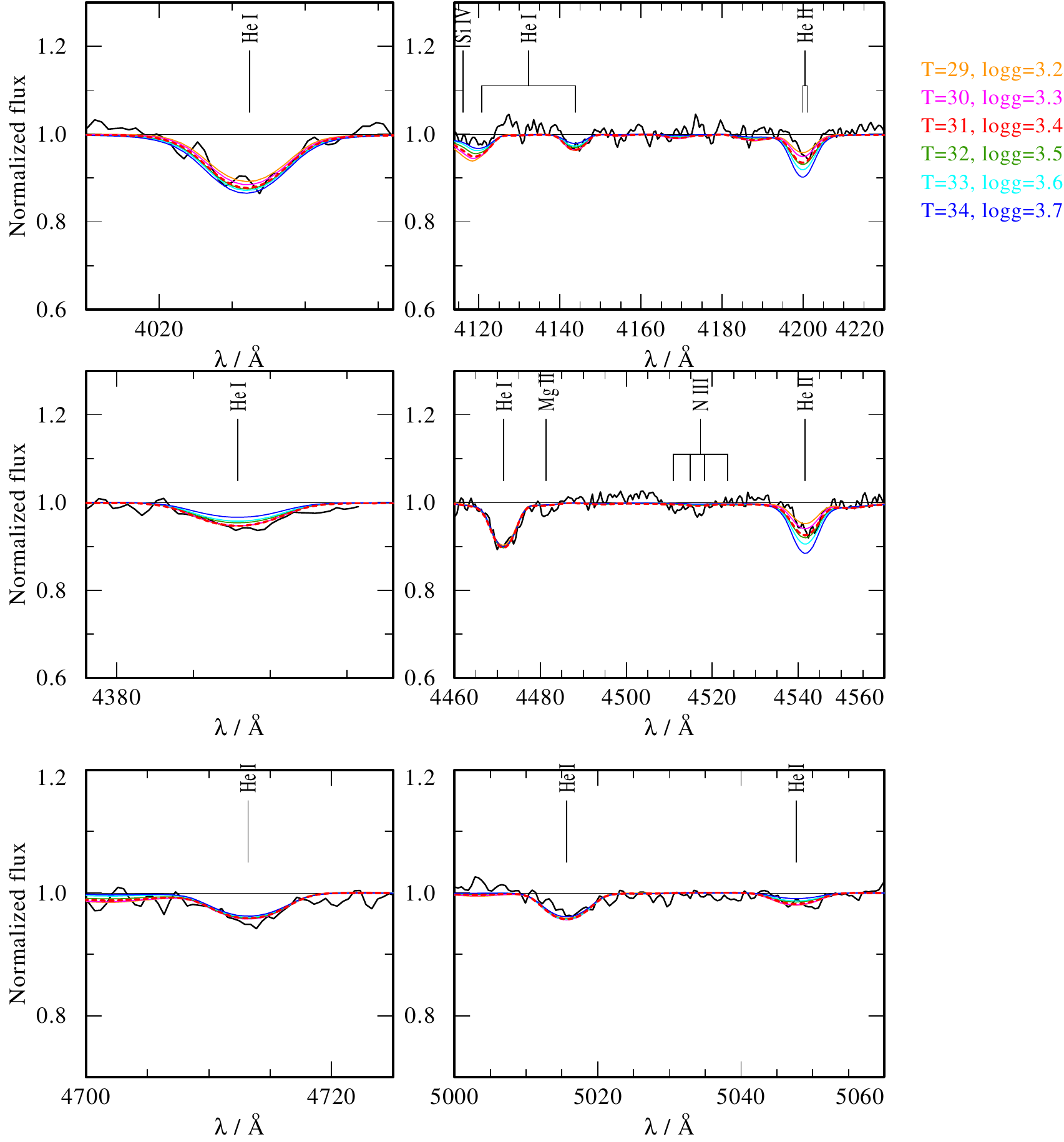} 
\end{figure*}

\end{document}

%% file: bibdefinitions.tex
\DeclareRobustCommand{\ion}[2]{%
	\relax\ifmmode
	\ifx\testbx\f@series
	{\mathbf{#1\,\mathsc{#2}}}\else
	{\mathrm{#1\,\mathsc{#2}}}\fi
	\else\textup{#1\,{\mdseries\textsc{#2}}}%
	\fi}
\def\degr{\hbox{$^\circ$}}

\newcommand{\revise}{ }

\newcommand{\mx}{\mbox{M33\,X-7}}
\newcommand{\xmm}{{\em XMM-Newton}}

\newcommand{\hst}{{\em HST}}

\newcommand{\hii}{\ion{H}{ii}\ }
\newcommand{\hei}{ \ion{He}{i}\ }
\newcommand{\heii}{ \ion{He}{ii}\ }

\newcommand{\flux}{erg\,cm$^{-2}$\,s$^{-1}$}

\def\apjl{ApJ}%
\def\apjs{ApJS}%
\def\ao{Appl.~Opt.}%
\def\aap{A\&A}%